\documentclass{aa}
\usepackage{amssymb}
\usepackage{multirow}
\usepackage{rotating}
\usepackage{ctable}
\usepackage{morefloats}
\usepackage{graphicx,epsfig,fancyhdr,psfig,rotating,amsmath,epsf,txfonts,natbib,epstopdf,multirow,appendix}
\usepackage[colorlinks, citecolor=blue]{hyperref}

\def \halpha {H$\alpha$}
\def \kms {{ \rm km\;s$^{-1}$}}

\def \arcsec {$^{''}$}
\graphicspath{{figs/}}
\begin{document}
\title{H$\alpha$ spectroscopy and  multiwavelength imaging of a solar flare caused by filament eruption}

\author{Z. Huang
	\inst{1,2}
\and M.\,S. Madjarska
	\inst {2}
\and
     K. Koleva
    \inst{3}
	\and
J.\,G. Doyle
\inst{2}
\and
P. Duchlev
\inst{3}
\and
 M. Dechev
\inst{3}
\and
K. Reardon
\inst{4}}
\offprints{zhu@arm.ac.uk}

\institute{
Shandong Provincial Key Laboratory of Optical Astronomy and Solar-Terrestrial Environment, School of
Space Science and Physics, Shandong University (Weihai), 264209 Weihai, Shandong, China
\and
Armagh Observatory, College Hill, Armagh BT61 9DG, N. Ireland
 \and
Institute of Astronomy and National Astronomical Observatory,
BAS, 72 Tsarigradsko Chaussee blvd., 1784 Sofia, Bulgaria
 \and
National Solar Observatory, Sacramento Peak, P.O. Box 62, Sunspot, NM, USA, 88349}

 \date{Received date, accepted date}

\abstract
{We study a sequence of eruptive events including filament eruption, a GOES C4.3 flare, and a coronal mass ejection.}
{We aim to identify the possible trigger(s) and precursor(s) of the filament destabilisation, investigate flare kernel characteristics, flare ribbons/kernels  formation and evolution, study the interrelation of the filament-eruption/flare/coronal-mass-ejection phenomena as part of the integral active-region magnetic field configuration, and determine \halpha\ line profile evolution during the eruptive phenomena.}
{Multi-instrument observations are analysed including \halpha\ line profiles, speckle images at \halpha\,--\,0.8\,\AA\
and \halpha\,+\,0.8\,\AA\ from IBIS at DST/NSO, EUV images and magnetograms from the
SDO, coronagraph images from STEREO, and
the X-ray flux observations from FERMI and GOES.}
{We establish that the filament destabilisation and eruption are the main triggers for the flaring activity. A surge-like
event with a circular ribbon in one of  the filament footpoints is determined as the possible trigger of the filament
destabilisation. Plasma draining in this  footpoint is identified as the precursor for the filament eruption.
A magnetic flux emergence prior to the filament destabilisation followed by a high rate of flux
cancellation of $1.34\times10^{16}$\,Mx\,s$^{-1}$ is found during the flare activity. The flare X-ray lightcurves
reveal three phases that are found to be associated with three different ribbons occurring consecutively. A kernel from each ribbon is selected and analysed. The kernel lightcurves and \halpha\ line profiles reveal that the emission increase in the line centre is stronger than that in the line wings. A delay of around 5--6 mins is found between the increase in the line centre and the
occurrence of  red asymmetry. Only red asymmetry is observed in the ribbons during
the impulsive phases. Blue asymmetry is only associated with the dynamic filament.}
{}

\keywords{Sun: activity - Sun: flare - Sun: filaments - Line: profiles - Methods: imaging and spectroscopy}
\titlerunning{H$\alpha$ spectroscopy and EUV imaging of a C4.3 flare}

\maketitle

\section{Introduction}
\label{sect_intro}

Solar flares are powerful solar phenomena that are believed to be driven by magnetic reconnection
resulting in plasma heating and particle acceleration. They can be
observed as emission enhancements across the entire electromagnetic spectrum, from radio to $\gamma$-ray wavelengths. Flares
are considered as phenomena initiated in the corona since radio and hard X-ray emission at flaring sites
were discovered  \citep[and the references therein]{lrsp-2011-6}. For decades, the chromospheric response to flares has been investigated by using \halpha\ filtergrams. Flaring sites observed in \halpha\ show spectacular phenomena such as filament (prominence) eruptions and flare ribbons (bright regions in the chromosphere along the magnetic neutral line); \halpha\ kernels, which are very bright and compact \halpha\ emission sources embedded in flare ribbons, are also
common features appearing during a flare. They are believed to be the locations of high-energetic particle precipitation.
More details about solar flares can be found in some reviews
\citep[e.g. ][etc.]{2007ASPC..368..365H,lrsp-2008-1,lrsp-2011-6,2011SSRv..159...19F}.

\par
Although flares have been observed in chromospheric temperature since the \halpha\ filter was invented in the 1930s,
the precise mechanism(s) by which energy release in the corona drives chromospheric emission bursts, called  ribbons or kernels,  has not been
well established.  A two-dimensional magnetic reconnection model called CSHKP
\citep{1964NASSP..50..451C,1966Natur.211..695S,1974SoPh...34..323H,1976SoPh...50...85K}, suggests that the plasma surrounding a
null point in the corona is heated such that high coronal pressure, thermal conduction, and non-thermal particles
(mostly electrons) can efficiently carry energy from the magnetic reconnection site in the corona to the lower
solar atmosphere along the magnetic field lines \citep{1996ApJ...466.1054M}. Thermal radiation from soft X-rays,
EUV, and UV can also contribute to this process, but this contribution was found to be very small\,\citep{2005ApJ...630..573A}.
Other more recent works have raised questions about the viability of this mechanism in the light of recent 
observations \citep{2008ApJ...675.1645F} and suggested Alfv\'en wave propagation as an alternate energy
transport mechanism from the corona to chromosphere during flares \citep{2013ApJ...765...81R}.


\begin{figure*}[!ht]
\centering
\includegraphics[width=17cm]{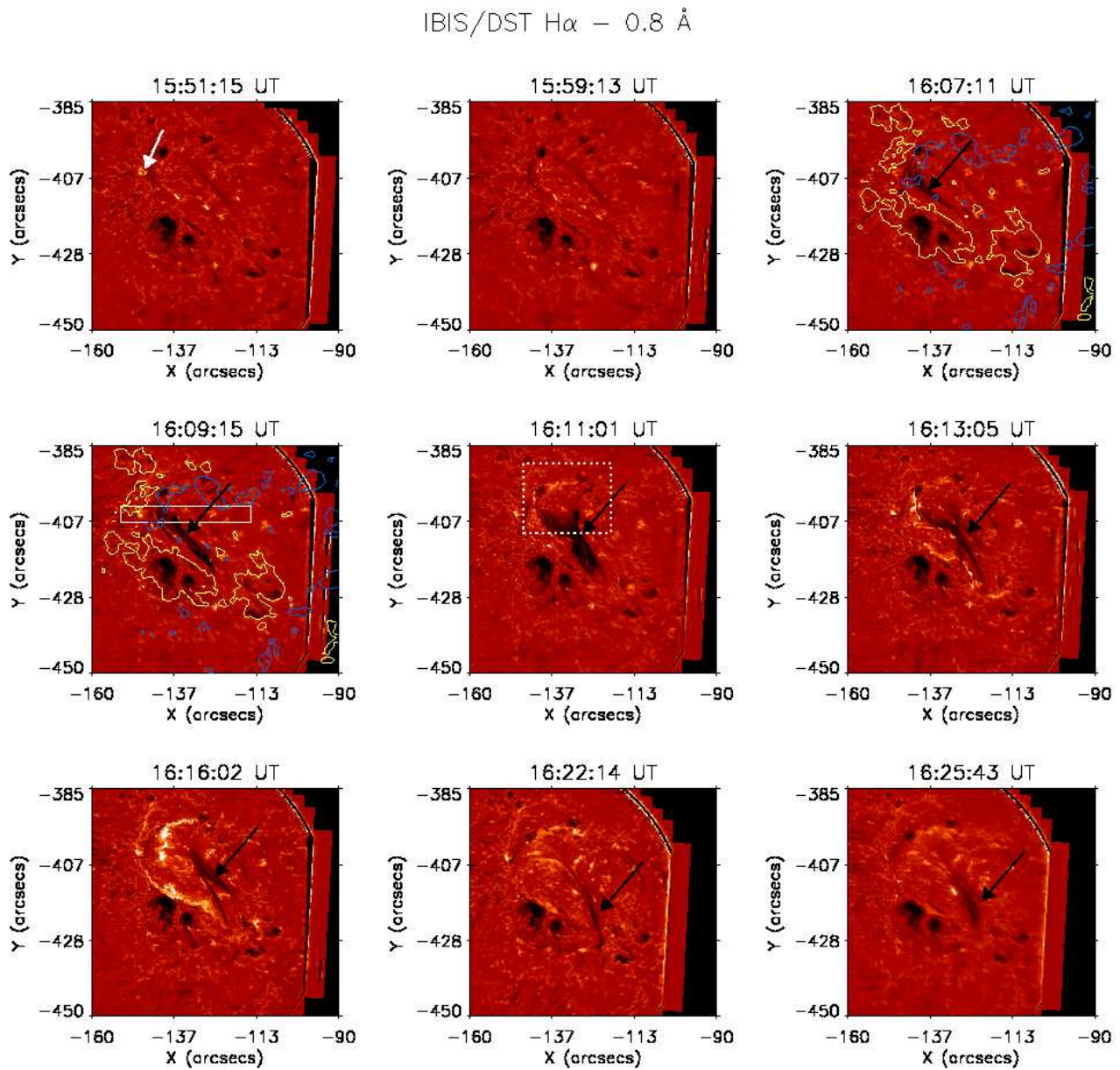}
\caption{The filament eruption and the flare ribbon seen in H$\alpha$ -- 0.8~\AA. The white arrow at 15:51~UT marks the ribbon associated with the surge-like event in the footpoint of the filament. The black arrows indicate
the rising twisted flux-ropes of the filament. The white box at 16:09~UT denotes the region used to produce
the slice-time images in Fig.~\ref{fig4}. The dotted lines at 16:11\,UT denote a region that is enlarged in
Fig.~\ref{fig8}. Contour plots of HMI magnetogram are shown in images at 16:07:11\,UT and 16:09:15\,UT (yellow: -300\,Mx\,cm$^{-2}$, blue: 300\,Mx\,cm$^{-2}$).}
\label{fig1}
\end{figure*}

\begin{figure*}[!ht]
\centering
\includegraphics[width=17cm]{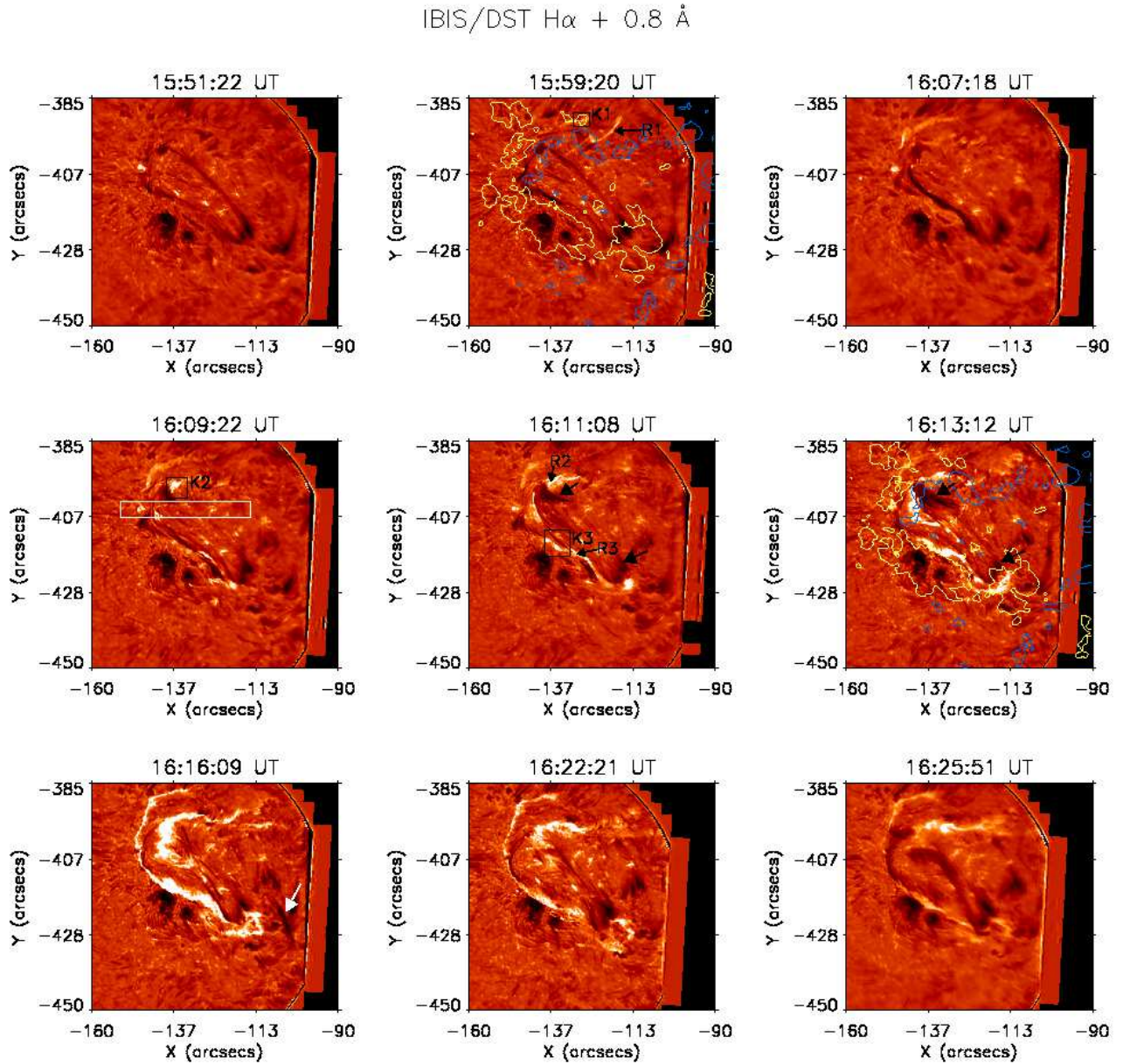}
\caption{The filament eruption and the flare ribbon seen in \halpha + 0.8~\AA. Larger black arrows indicate the
footpoints of the filament where down-flows are observed (at 16:11~UT). Three kernels (K1, K2, and K3) are noted on the
images at 15:59\,UT, 16:09\,UT, and 16:11\,UT, respectively. The ribbons (R1, R2, and R2) are shown with arrows in the images at 15:59~UT and 16:11~UT. The white box in the image at 16:09\,UT indicates
the region used to produce the time-slice plot in Fig.~\ref{fig4}. The white arrow at 16:16:09~UT
marks the top of the falling back filament ropes (or part of it). Contour plots of HMI magnetogram are shown in images at 16:07:11\,UT and 16:09:15\,UT (yellow: -300\,Mx\,cm$^{-2}$, blue: 300\,Mx\,cm$^{-2}$).}
\label{fig2}
\end{figure*}


\begin{figure*}[!ht]
\centering
\includegraphics[width=17cm]{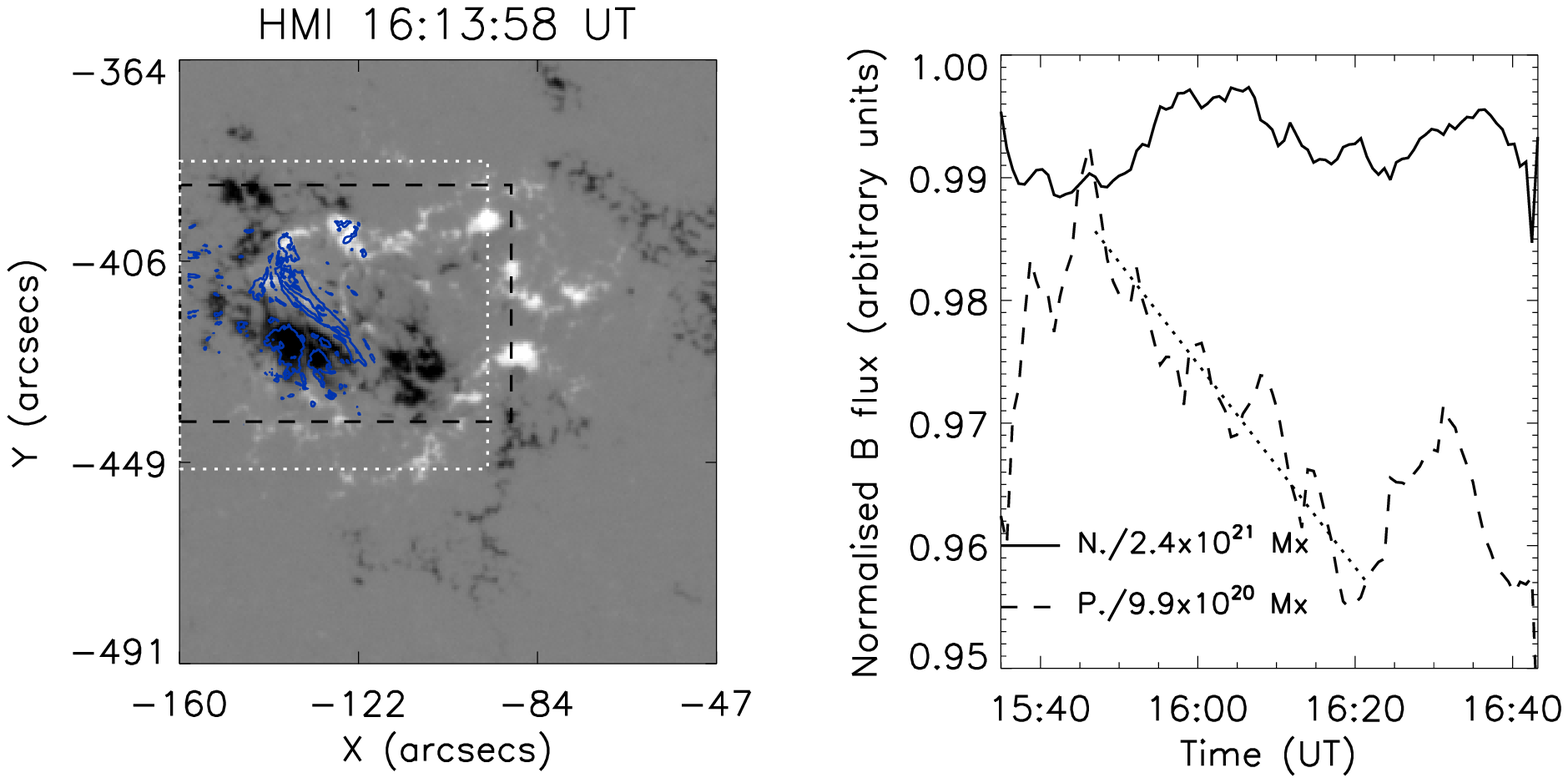}
\caption{\textbf{Left:} HMI longitudinal magnetogram of the flare scaled from -1 kG to 1 kG. The region
enclosed with dotted lines is the IBIS field of view. The overplotted contours (blue lines) outline the
filament and two of the sunspots in the flaring region. \textbf{Right:} lightcurves of the normalised
(to the maximum) magnetic flux  from the region shown with black dashed line in the left panel.  The solid line
indicates the variation of the negative flux while the dashed line is the positive one. The cancellation rate
of the positive flux is obtained from the linear fit of the positive flux lightcurve shown here with a dotted
straight line.}
\label{fig3}
\end{figure*}

\par
Although \halpha\ filtergrams provide a wealth of information on the dynamic morphological evolution of the flare in the
chromosphere\,\citep[][and references therein]{2007ASPC..368..365H}, full \halpha\ line profiles have powerful diagnostic
potential for understanding the physical mechanism driving solar flares. Based
on a static model, \citet{1984ApJ...282..296C} calculated \halpha\ profiles of flare chromospheres produced by
different mechanisms (see the previous paragraph). They found that central reversal effects on the profiles depend on
the heating rate by non-thermal electrons, while increasing thermal conduction reduces the width and total intensity
of the profiles, thus lessening the contribution to the flare \halpha\ enhancement. They pointed out that a high
coronal pressure can dramatically enhance the width and total intensity of the \halpha\ line and only a high value of
non-thermal electron flux can produce \halpha\ profiles with non-Gaussian broad wings. \citet{1993A&A...274..917F}
also considered the non-thermal excitation and ionisation effect in the flare and suggested that bombardment of
the non-thermal electrons in the chromosphere can also significantly strengthen and broaden \halpha\ profiles. However,
the detection of the effect of non-thermal electron beams on \halpha\ profiles is still under debate.
\citet{2009A&A...499..923K} concluded from their simulations that the non-thermal electron beams generally result in
emission enhancement although they can also affect the line wings. It is generally accepted that different heating
mechanisms most probably work together \citep[e.g.][]{2005A&A...435..743S,2006ApJ...653..733C}.

\par
Enhanced emission in the red wing of the \halpha\ line, or a line-wing bisector that shifts more towards the red, are often observed 
in flare kernels. This feature is called  a ``red asymmetry'' \citep[][and references therein]{svestka1976}. Red
asymmetry is considered to be a signature of downward-moving chromospheric material which results from cooling and condensation of the chromospheric material 
previously heated during ``chromospheric
evaporation''. The term chromospheric evaporation refers to the process of the dense, cold chromospheric
plasma being heated to tens of million degrees by the energy released during the solar flare in the corona, 
and its subsequent expansion up into the corona
 \citep{1988ApJ...324..582Z,1990ApJ...348..333C,1990ApJ...363..318C}. Various authors
\citep[e.g.][]{1984SoPh...93..105I} have found that  \halpha\ kernels exhibiting red asymmetry are very small ($<$ 1\arcsec) and
change position continuously.  The lifetime of the red asymmetry is shorter than that
of the kernel. It has been reported that red asymmetry can be seen in many parts of the flare ribbons in
most flares\,\citep{1983SoPh...83...15T,1984SoPh...93..105I,1990ApJ...363..318C}. Recently, \citet{2012PASJ...64...20A}
found that the red asymmetry appears all over the flare ribbons with the strongest red asymmetry 
located on the outer narrow edges of the flare ribbons (with a width of 1.5\arcsec -- 3\arcsec) 
where the strongest energy release occurs.
\citet{deng2012} found that red asymmetry is observed only in the core area of the flare ribbons during the
impulsive phase.

\par
A different (or opposite) phenomenon, ``blue asymmetry'',  is also found in the early stages of
the impulsive phase \citep[e.g.][etc]{1962BAICz..13...37S,1983SoPh...83...15T,1990ApJ...363..318C,1992SoPh..141...91G,1994SoPh..149..195J}. However, observations of blue asymmetry are somewhat controversial and different mechanisms have been proposed
\citep{1994SoPh..152..393H}. They have been associated with chromospheric evaporation \citep{1974SoPh...34..323H} and
electron beam heating with return currents \citep{1994SoPh..152..393H}.  \citet{1994SoPh..149..195J} argue that blue asymmetry is an observational effect from multiple evolving loops. \citet{1984SoPh...93..105I} found  a momentum balance between the downward motion of the
chromospheric condensation and the chromospheric evaporation, thus confirming  that the red asymmetry is a counteraction
of the chromospheric evaporation. In some cases, both red and blue asymmetry can be
found in the same flare \citep{1990ApJ...363..318C}. These authors found blue asymmetry  associated with eruptive and
untwisting filaments in one case, and suggested that two other cases are related to unresolved filament activation occurring at a
very small spatial scale.

\par
Eruptive prominences (EPs) or filaments are frequently associated with coronal
mass ejections (CMEs) and flares \citep{1995ASSL..199.....T}. All three phenomena occur in the same
large-scale magnetic field configuration, but the precise relationship between these phenomena  has not yet been clearly
established. Hence, an investigation of the complex trio phenomenon (filament eruption/flare/CME) observed at
high-spatial and temporal resolution from the chromosphere to the corona is essential for understanding the
physical mechanism(s) driving each of these events.  

Triggers of prominence eruption are under debate, with various mechanisms being discussed and modelled during the
past few decades \citep[see review by][and the references therein]{2011LRSP....8....1C}. It has been suggested that new emerging magnetic
flux (which reconnects with the preexisting field) in the vicinity of a filament can trigger destabilisation of a
filament \citep{1995JGR...100.3355F}. 
Indeed, \citet{1998SoPh..182..145C} observed the emergence of flux with signatures of twist which injected that twisted magnetic field into the filament, leading to its destabilisation and eruption. The precursors of filament eruptions can be viewed as the key to understanding the physical mechanisms responsible for the
destabilisation and eruptions. \citet{1980SoPh...68..217M} reviews the primary signatures of the
pre-flare phase of filament eruptions among which are: increased absorption in the blue/red wing and line centre
as early as 30 min prior to the eruption, transition from absorption to emission of the prominence plasma in
transition region and coronal lines, and complete or partial disappearance in \halpha\ suggesting a heating process.
Many precursors have been found by different authors, for example, plasma drainage from the prominence to the chromosphere \citep{1995ASSL..199.....T,  2000ApJ...537..503G}, heating of filament plasma before \citep{2004SoPh..223...95C, 2006ApJ...645.1525K} and during eruptions
\citep{2001SoPh..202..293E, 2008ApJ...673..611K}, a cavity size increase \citep{2006ApJ...641..590G}, darkening
or brightening of parts of the filament body or its close vicinity in high temperature EUV lines
\citep{2006ApJ...653..719A}, and an X-ray brightening in the vicinity of the filament \citep{1985SoPh...97..387H}.


\begin{figure}[!ht]
\centering
\resizebox{\hsize}{!}{\includegraphics[clip,trim=0.5cm 0.5cm 0.5cm 1cm]{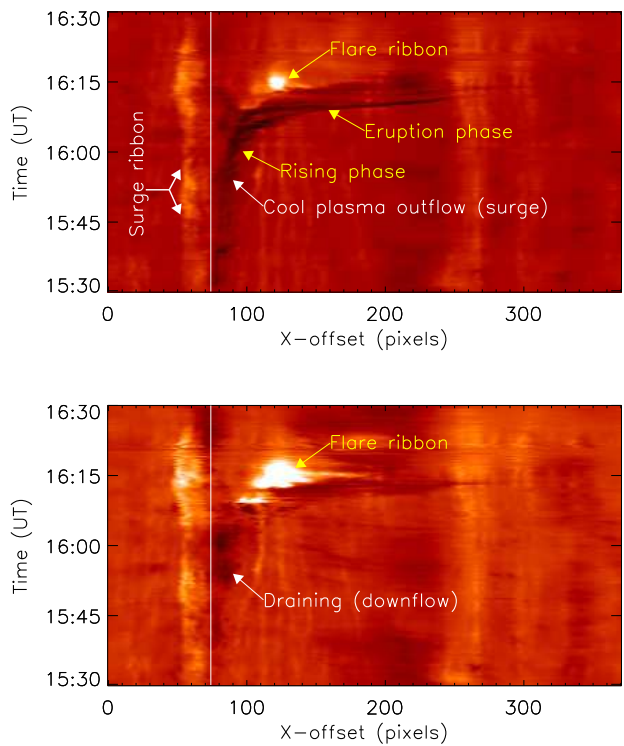}}
\caption{\textbf {Top:} Time-slice image in H$\alpha$ -- 0.8~\AA\ showing the ribbon associated with the surge-like
event, the rising filament, and its eruption as a dark feature. \textbf{Bottom:} \halpha\ + 0.8\,\AA\ time-slice
image showing plasma draining in the filament footpoint. The arrows indicate all identified phenomena, i.e.
surge-like down-flow (draining of cool material) and the flare's ribbon. The slice was taken at the location
indicated with a white box in Figs.~\ref{fig1} and \ref{fig2}.}
\label{fig4}
\end{figure}


\begin{figure}[!h]
\centering
\resizebox{\hsize}{!}{\includegraphics{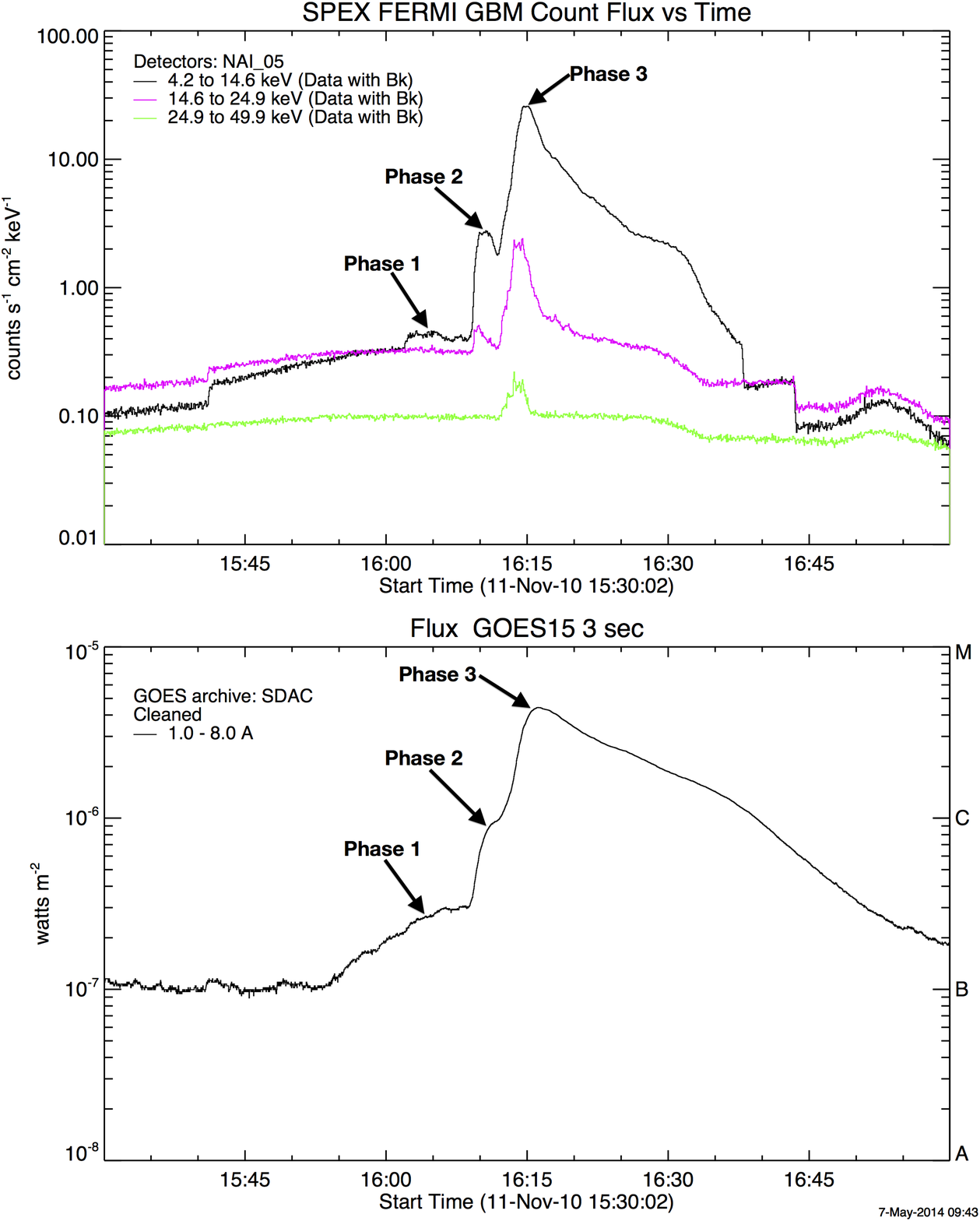}}
\caption{The solar X-ray flux measured by FERMI (top) and GOES (bottom). The arrows indicate the three impulsive peaks.}
\label{fig5}
\end{figure}


\begin{figure}[!ht]
\centering
\resizebox{\hsize}{!}{\includegraphics[clip,trim=0.5cm 0.3cm 0.5cm 0.5cm]{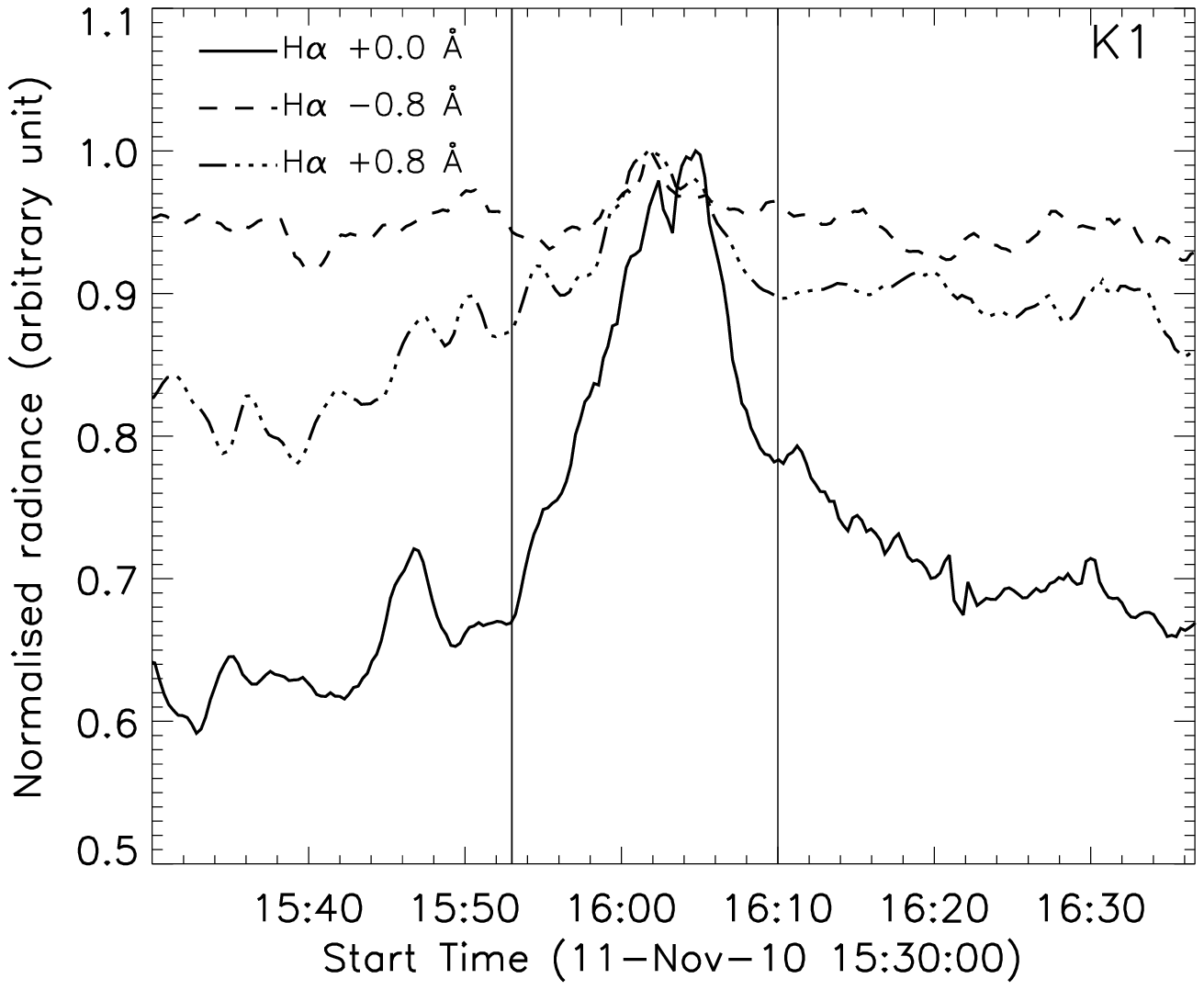}}
\resizebox{\hsize}{!}{\includegraphics[clip,trim=0.5cm 0.3cm 0.5cm 0.5cm]{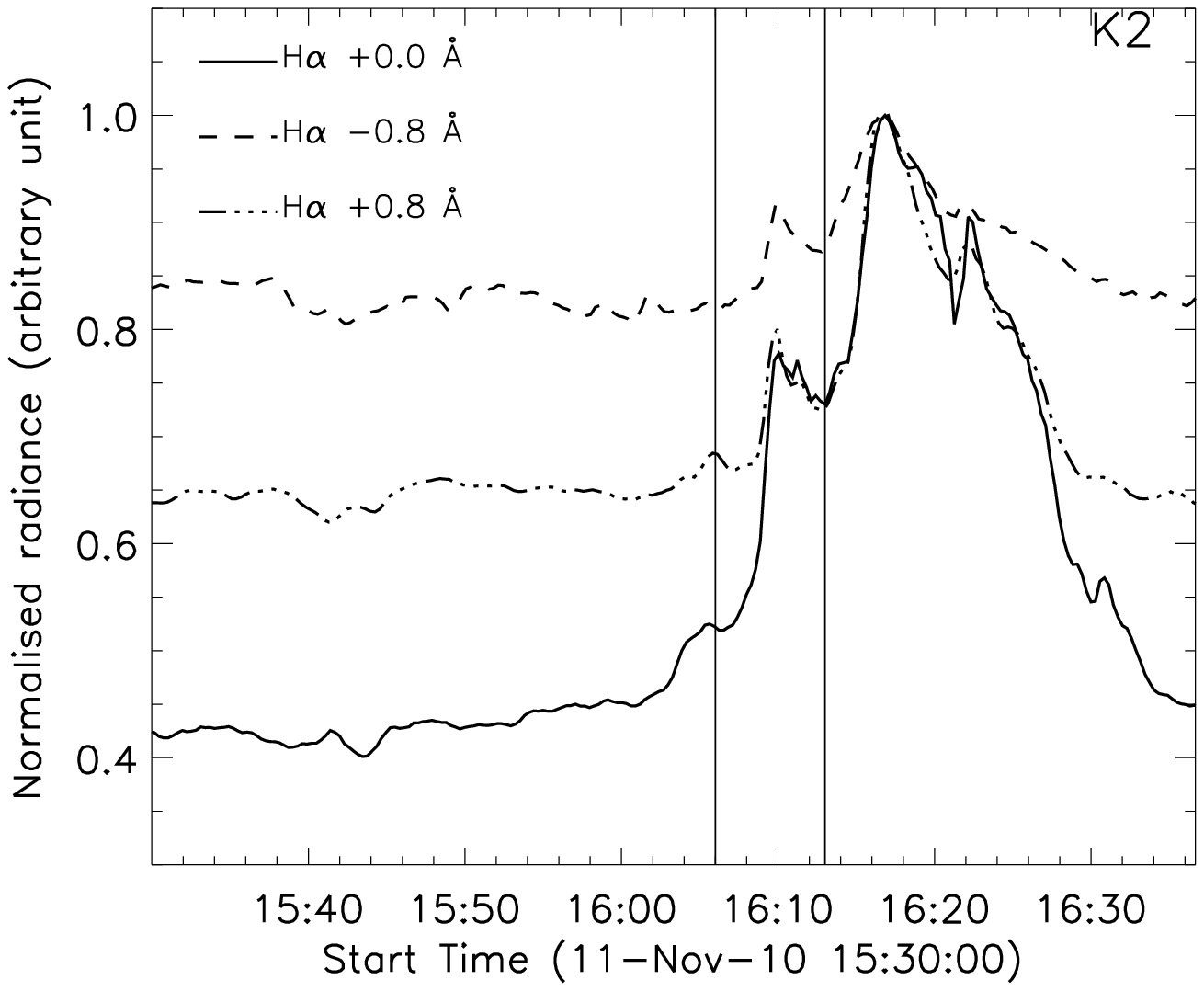}}
\resizebox{\hsize}{!}{\includegraphics[clip,trim=0.5cm 0.3cm 0.5cm 0.5cm]{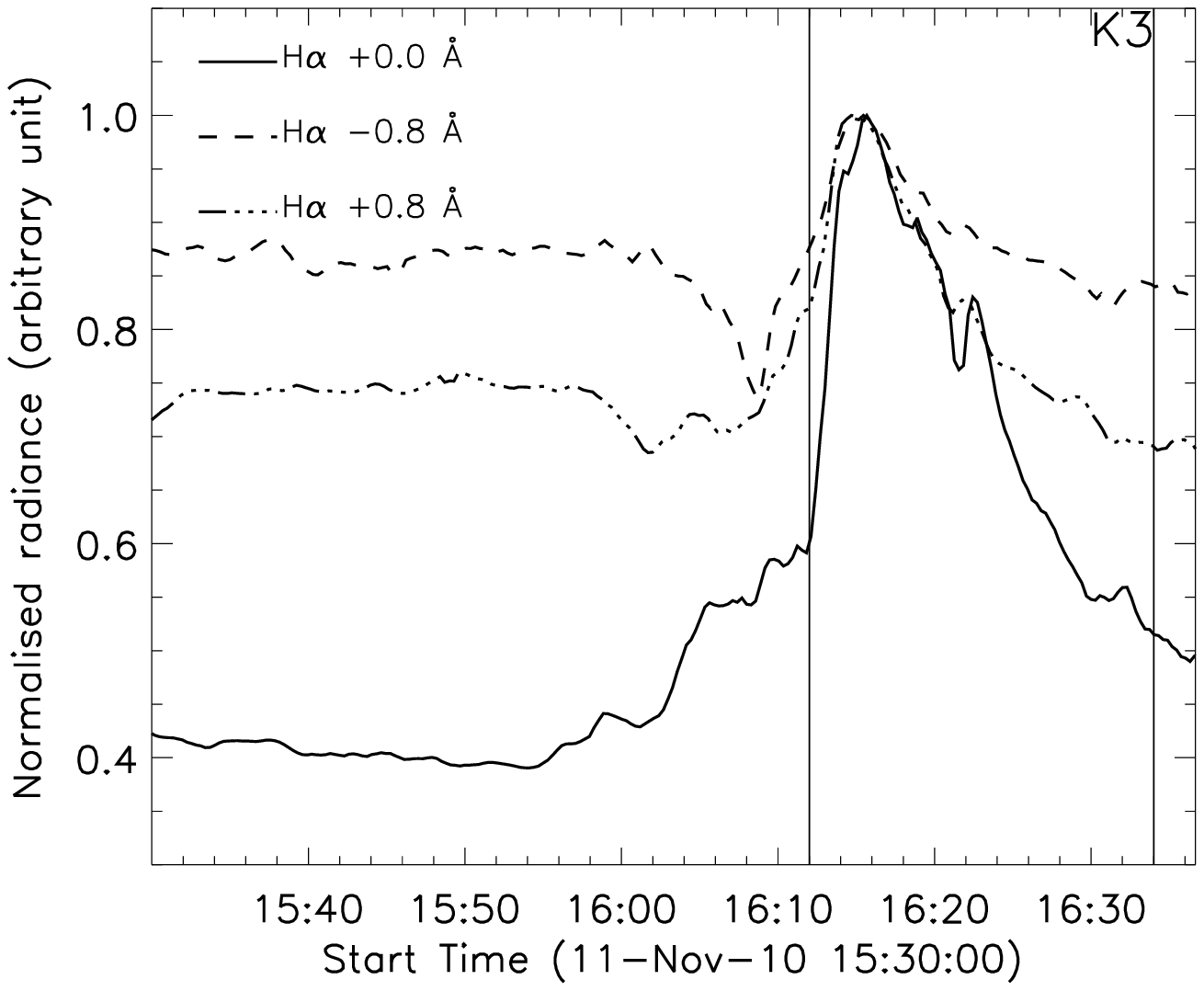}}
\caption{Lightcurves of the flare kernels outlined with boxes in Figure~\ref{fig2} in the \halpha\ line
centre, \halpha\ -- 0.8\,\AA, and \halpha\ + 0.8\,\AA.  The vertical lines indicate the onset time of each kernel occurrence. Top: K1; middle: K2; bottom: K3.}
\label{fig6}
\end{figure}

\begin{figure}[!ht]
\resizebox{\hsize}{!}{\includegraphics[clip,trim=0.5cm 0cm 0.5cm 0cm]{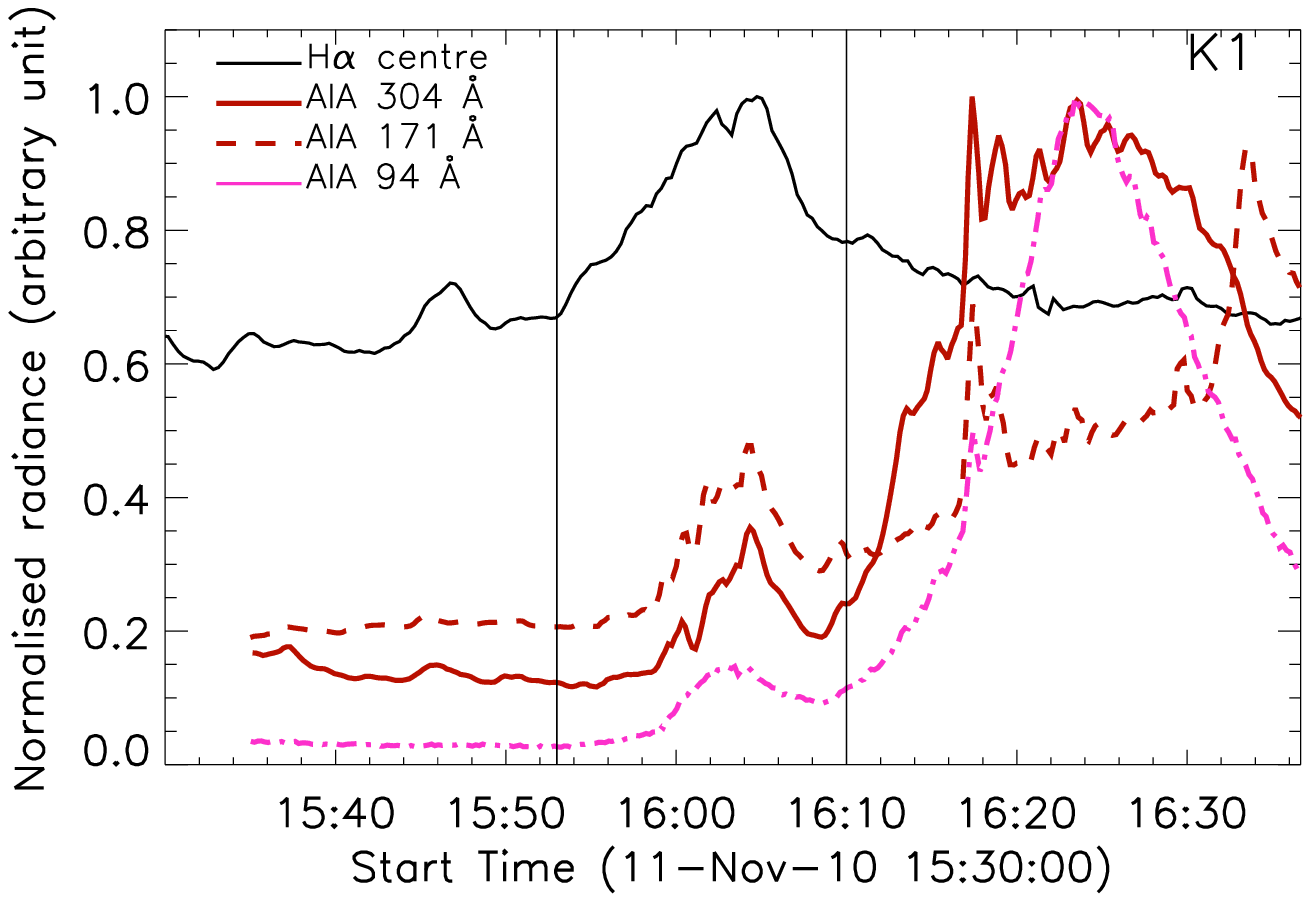}}
\resizebox{\hsize}{!}{\includegraphics[clip,trim=0.5cm 0cm 0.5cm 0cm]{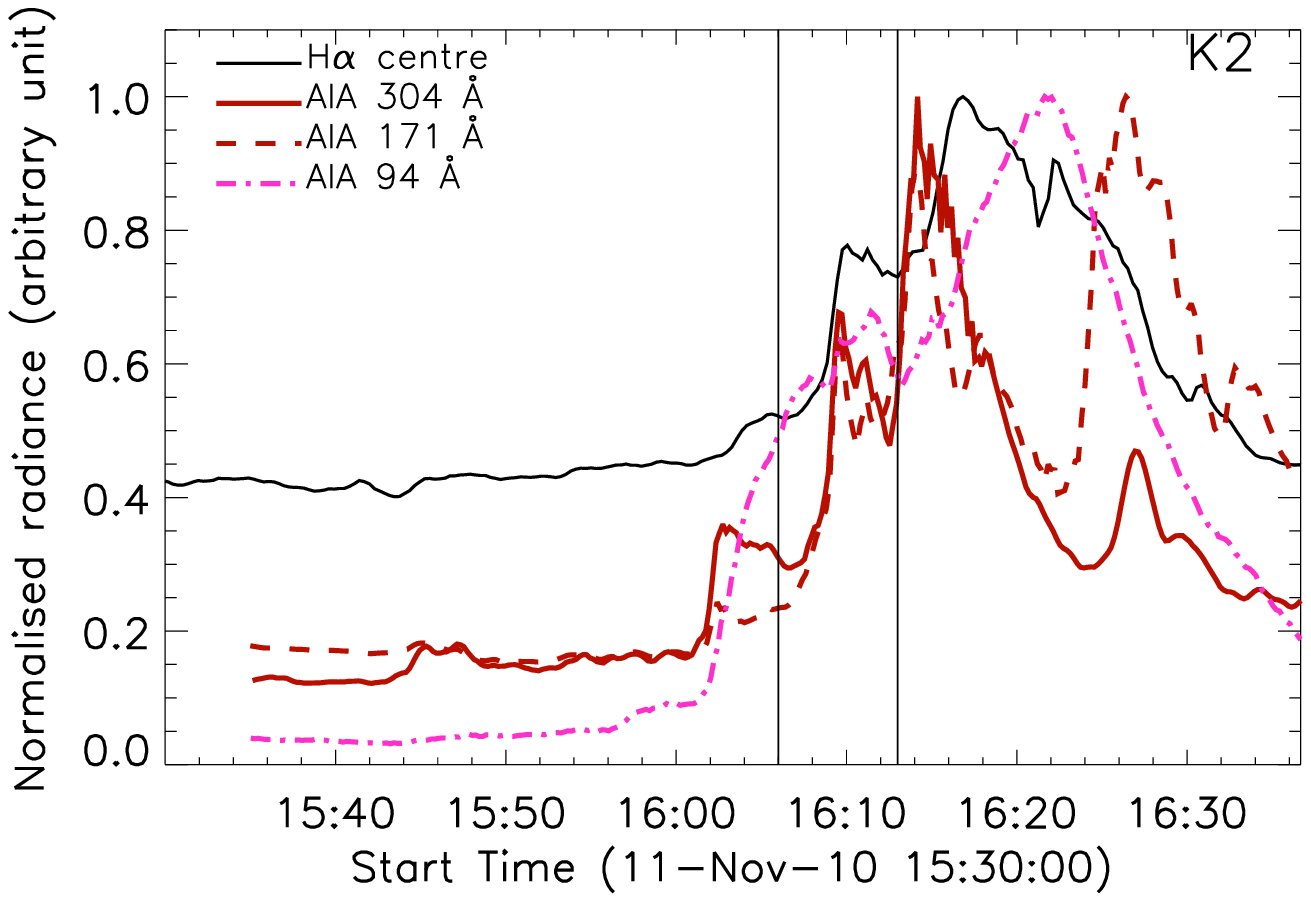}}
\resizebox{\hsize}{!}{\includegraphics[clip,trim=0.5cm 0cm 0.5cm 0cm]{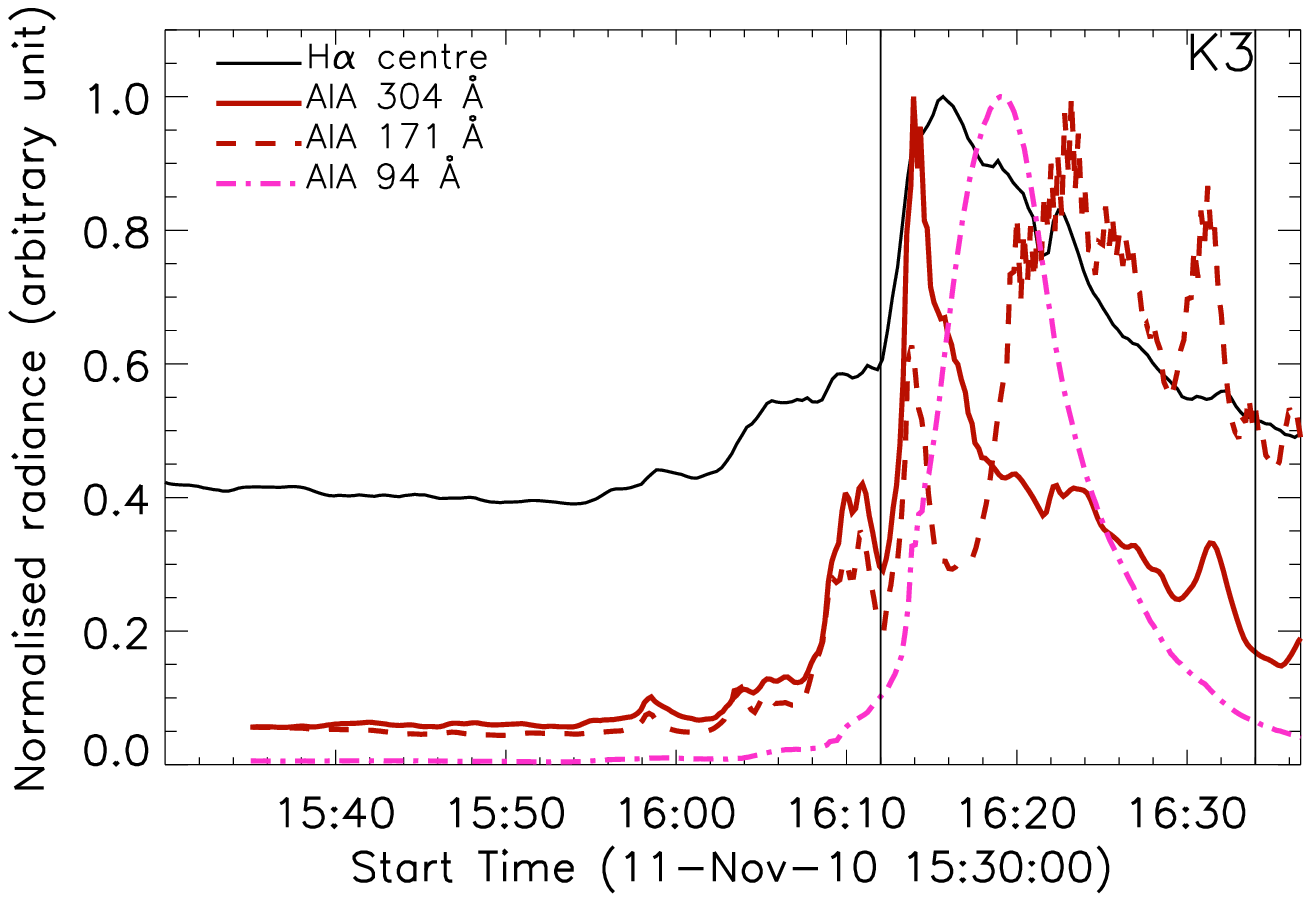}}
\caption{Lightcurves of the flare kernels outlined with boxes in Fig.~\ref{fig2} in the AIA 304\,\AA, 171\,\AA,
and 94\,\AA\ channels, and the corresponding \halpha\ line centre lightcurves are overplotted for reference.
The vertical lines indicate the onset time of each kernel occurrence. Top: K1; middle: K2; bottom: K3.}
\label{fig7}
\end{figure}

There have been numerous reports on solar flares and associated processes like eruptive filaments and CMEs.
Here, we report the observations of an active region near disk centre from the chromosphere (spectroscopically and imaging) through the
transition region and corona (imaging). We make an in-depth spectroscopic and imaging analysis of the complex
filament eruption/flare/CME phenomenon using space- and ground-based observations with largely unprecedented temporal and spatial resolution covering the entire solar atmosphere from the chromosphere to the corona. We study the trigger and filament-eruption precursors, analyse the \halpha\ profile searching for the mechanism producing chromospheric flare emission bursts, investigate the time evolution of red- and blue asymmetries and the phenomena causing their occurrence, analyse the magnetic flux evolution, and examine the formation of the coronal mass ejection with respect to the filament/flare events as they unfold. In Section~2 of the article we describe the observations. Section 3 presents  the results and discussion of the filament eruption (Section~3.1),
flare evolution  (Section~3.2), the \halpha\ spectroscopy study (Section~3.3), magnetic field evolution (Section~3.4),
and the CME association (Section~3.5). The obtained results are summarised in Section~4.

\section{Observations}
Our dedicated Hinode/Dunn Solar
Telescope (DST) observing campaign took place from 2010 November 10 to 18. On November 11 we
targeted NOAA 11123,  a complex $\beta\gamma$ active region composed of eight sunspots, and observations were taken in the time interval from 15:14\,UT to 17:08\,UT.
The active region
produced five GOES C class flares with the strongest of GOES class C4.7.
Unfortunately, Hinode was not targeting the active
region, so we do not have crucial information from the three Hinode instruments. Nevertheless,
unique observations of this flare were taken with several instruments from ground- and
space-based observatories. We used data from the 
Interferometric BIdimensional Spectrometer~\citep[IBIS,][]{2006SoPh..236..415C}  mounted on the DST at the National Solar Observatory (NSO) at Sacramento Peak
which is equipped with a high-order adaptive optics system. The space-based instruments include
the Atmospheric Imaging Assembly (AIA) and the Helioseismic and Magnetic Imager (HMI) on board the Solar Dynamics
Observatory (SDO), the Extreme Ultraviolet Imager (EUVI) and COR1 inner coronagraph  on board the Solar Terrestrial
Relations Observatory (STEREO) Behind (B) and Ahead (A) spacecraft, the Gamma-ray Burst Monitor (GBM) on board FERMI and the Geostationary Operational Environmental Satellite
(GOES). A GOES C4.3 flare was registered by all instruments starting at 15:54\,UT together with
a filament eruption followed by a coronal mass ejection recorded by LASCO/SoHO and
COR1/COR2/SECCHI/STEREO A and B. A description of the instruments that were used together with details
on the obtained data and their reduction are given below.

\subsection{IBIS}
IBIS uses a pair of Fabry-Perot interferometers to record very narrow-band filtergrams in the
spectral range 5800--8600\,\AA\,\citep{2006SoPh..236..415C, 2008A&A...481..897R}.  It has a
95\arcsec diameter circle field of view and a pixel size of about 0.1\arcsec. Please note that we study the flare 
in a 65\arcsec x 65\arcsec area that was cut from the IBIS circular field of view. The observations
were taken in the H$\alpha$\,6562.8\,\AA\ line. A sequence of data segments was specially designed to
obtain both full spectral scan and seeing-free images at two selected wavelengths at the highest possible cadence. For each data segment,
IBIS first scanned the H$\alpha$ line from $-$1.4\,\AA\ to $+$1.4\,\AA\ centred on the line centre with a
0.2\,\AA\ step for a total of 15 points, followed by 50 images taken at a
fixed wavelength at 6562\,\AA\ (i.e. H$\alpha$ -- 0.8\,\AA), and then 50 images at
6563.6\,\AA\ (i.e. H$\alpha + $0.8\,\AA). We named the images taken at a fixed wavelength
``speckle stacks''. The frame rate was about 5 frames per second and the exposure time was 35\,ms.
The full sequence was repeated every 17\,s. Before the science observations, IBIS took
a dataset by scanning the H$\alpha$ line profile from H$\alpha$ -- 2.0\,\AA\ to H$\alpha$ + 2.0\,\AA\
with about 36\,m\AA\ per step, which was used for pre-filter correction. A dataset for
dark-current calibration was taken with the sunlight blocked, and a dataset for
flat-field correction was obtained at the same wavelengths as the line-scan data.


\par
The data reduction is different for line-scan and speckle stack data. For the line-scan data,
the calibration includes dark-current, flat-field, blue-shift, pre-filter, and destretching corrections
\citep[see e.g. ][for details]{2009A&A...503..577C}. For the speckle stack, the dark-current and
flat-field corrections are first performed, then each speckle stack (50 images) is used as
input to an imaging reconstruction package called the Kiepenheuer-Institut Speckle Interferometry
Package\,\citep[KISIP,][and references therein]{2008A&A...488..375W} to produce a single output
image. The KISIP package is designed to reconstruct solar speckle
interferometric data observed using an AO system \citep{2008A&A...488..375W}. The image
reconstruction improves the quality of the images that were blurred owing to the turbulent
atmospheric conditions. After this calibration process, each data segment consists of a  line-scan image,
one image at H$\alpha$ -- 0.8\,\AA, and one image at H$\alpha$ + 0.8\,\AA\ at a cadence of 17\,s.

\begin{table*}
\centering
\caption{Timeline of the filament-eruption/flare/CME activities.}
\label{table1}
\begin{tabular}{p{1.2cm} |p{3cm}| p{3cm} |p{3cm}| p{3cm}| p{2cm}}
\hline\hline
Time (UT) &Filament activities& \multicolumn{3}{c|}{Flare activities}&{CME}\\
\cline{3-5}
&&X-ray&\halpha&EUV&\\
\hline
15:45& Circular ribbon and surge-like activity&&&\\
\hline
15:53&Filament starts slowly rising &&\halpha\ line centre of K1 starts rising&\\
\hline
15:58&&&Red asymmetry in K1 starts&\\
\hline
15:59&Filament eruption starts&&&\\
\hline
16:02&&Peak of the 1st impulsive phase&Peak of K1 \halpha\  line centre intensity&Peaks of the K1 EUV lightcurves (LCs)\\
\hline
16:03&&& Line centre in K2 starts rising&\\
\hline
16:05&&& Line centre in K3 starts rising&\\
\hline
16:08&&&Red asymmetry  in K2 starts&\\
\hline
16:10&Full eruptive phase of the filament&Peak of the 2nd impulsive phase&Peak of K2 \halpha\ line centre intensity&Peaks of the K2 of AIA 304~\AA\ and 171~\AA\ LCs\\
\hline
16:11&&&Red asymmetry in K3 starts&\\
\hline
16:13&&&&Peak of the K2 AIA 94~\AA\ LC&\\
\hline
16:14 to 16:15&&Peak of the 3rd impulsive phase&Peak of K3 \halpha\ line centre intensity &Peaks of  the K3 AIA 304~\AA\ and 171~\AA\ LCs&First CME cloud in COR1\\
\hline
16:16&Falling back of filament material observed&&&\\
\hline
16:19&&&&Peak of the K3 AIA 94~\AA\ LC&\\
\hline
16:25 to 16:30&&&&&Second CME cloud in COR1\\
\hline
\end{tabular}
\label{table1}
\end{table*}

\subsection{AIA/HMI/SDO}
We analysed data from AIA~\citep{2012SoPh..275...17L} and HMI~\citep{2012SoPh..275..229S}
on board SDO ~\citep{2012SoPh..275....3P}. The AIA data used here were taken in the 1600\,\AA, 304\,\AA,
171\,\AA, 193\,\AA, and 94\,\AA\ passbands at 12~s cadence and exposure time that varies during the flare.
The response of these channels to different temperature plasma can be found in
\citet{2010A&A...521A..21O}. The HMI measures Doppler shifts, intensity, and vector magnetic field
using the Fe~{\sc i}  6173~\AA\ absorption line. For the present study
only the longitudinal magnetograms were analysed. The AIA images have a pixel size of 0.6\arcsec.
The HMI magnetograms have a pixel size of 0.505\arcsec, a cadence of 45~s, and  a
1$\sigma$ noise level
of 10~G \citep{2012SoPh..279..295L}.

\subsection{Fermi}
Data from GBM \citep{2009ApJ...702..791M} on board the $Fermi$
mission were also analysed for this study. The GBM  has  12 sodium iodide (NaI) detectors which cover the
energy range from 8~keV to 1~MeV and two bismuth germanate (BGO) detectors covering the range from
200~keV to 40~MeV. The detectors are positioned on the side of the spacecraft in order to
view the part of the sky not obscured by the Earth. In this paper we use data 
only from the  sun-ward oriented NaI detector labeled n5.
For this
flare only the CSPEC (Continuous Spectroscopy) data were available; CSPEC has 128 energy
channels that take data at  4.096~s nominal time resolution, which can change to 1.024~s (trigger mode)
if a sufficient count rate is reached.

\subsection{EUVI/COR1/STEREO}

We analysed observations from the EUVI \citep{2008SSRv..136...67H} and the COR1 inner coronagraph
on board STEREO-B and A \citep{2008SSRv..136....5K}. EUVI has a field of view
of 1.7 R$_{\odot}$ and observes in four spectral channels, He~{\sc ii} 304~\AA, Fe~{\sc ix/x}
171~\AA, Fe~{\sc  xii} 195~\AA, and Fe~{\sc xiv}~284~\AA, that cover the 0.1 MK to 20 MK
temperature range. The EUVI detectors have a 2048$\times$2048 pixel$^2$ field-of-view with a pixel
size of 1.6\arcsec. In the present study, we used images in the He~{\sc ii} 304~\AA\ and
Fe~{\sc ix/x}~193~\AA\ channels with an average cadence of 10 min and 2 min, respectively.
COR1 is a classic Lyot, internally occulting, refractive coronagraph with a field of view from 1.3 to 4 solar radii. The separation
angle between the two satellites (A) and (B) on 2010 November 11 was 167.412$^{\circ}$,
and the separation angles with Earth were +84.5$^{\circ}$ for A and -82.9$^{\circ}$ for B.

\vspace{0.5cm}
The co-alignment of all data is made by aligning images covering similar temperature ranges. The
white-light images taken by SDO/AIA and DST were used as references when it was required.

\begin{figure*}[!ht]
\resizebox{\hsize}{!}{\includegraphics[clip,trim=0.5cm 1cm 0.5cm 1.5cm]{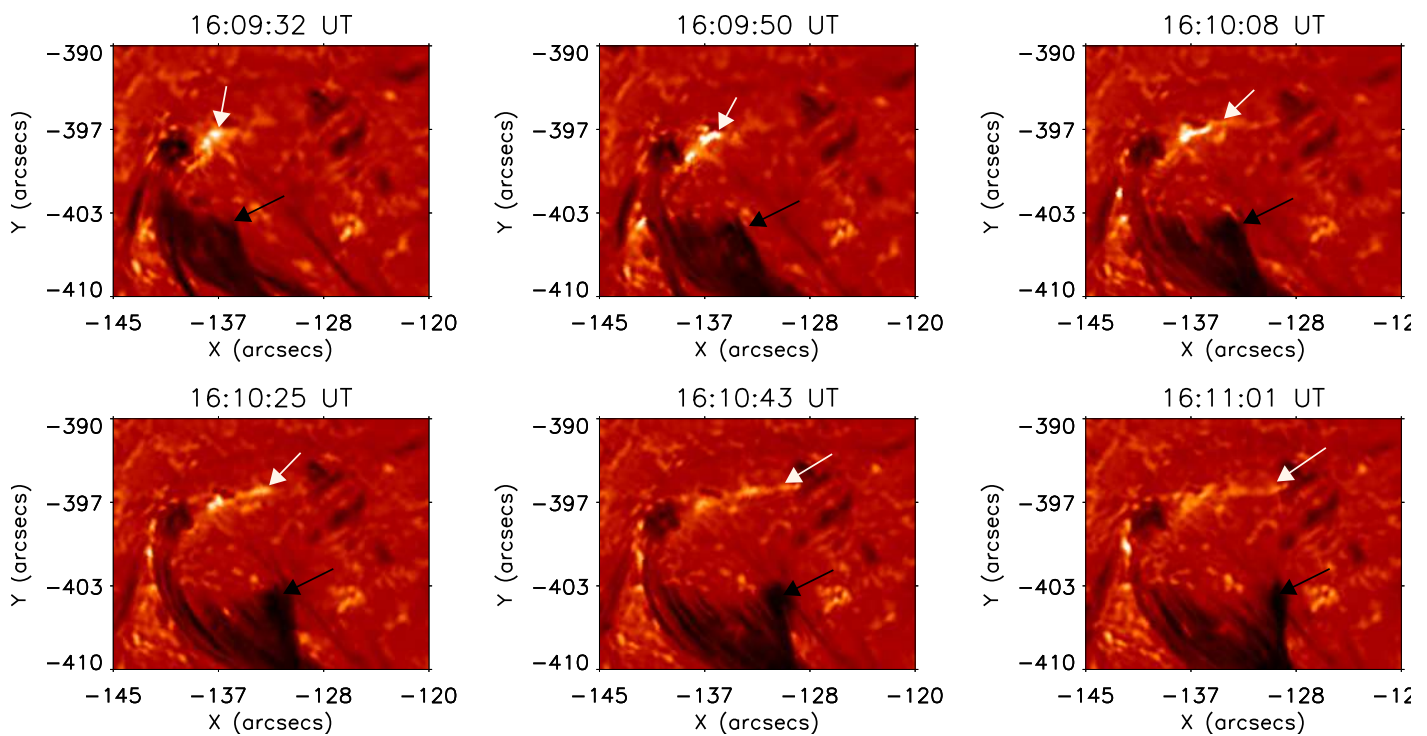}}
\caption{\halpha\ blue wing images enlarged from the region denoted as dotted lines in Fig.~\ref{fig1} at 16:11\,UT.
The arrows indicate the location in each image of the rising filament\,(black) and simultaneously forming ribbon\,(white) in its upper footpoint.}
\label{fig8}
\end{figure*}


\begin{figure*}[!ht]
\centering
\includegraphics[width=17cm]{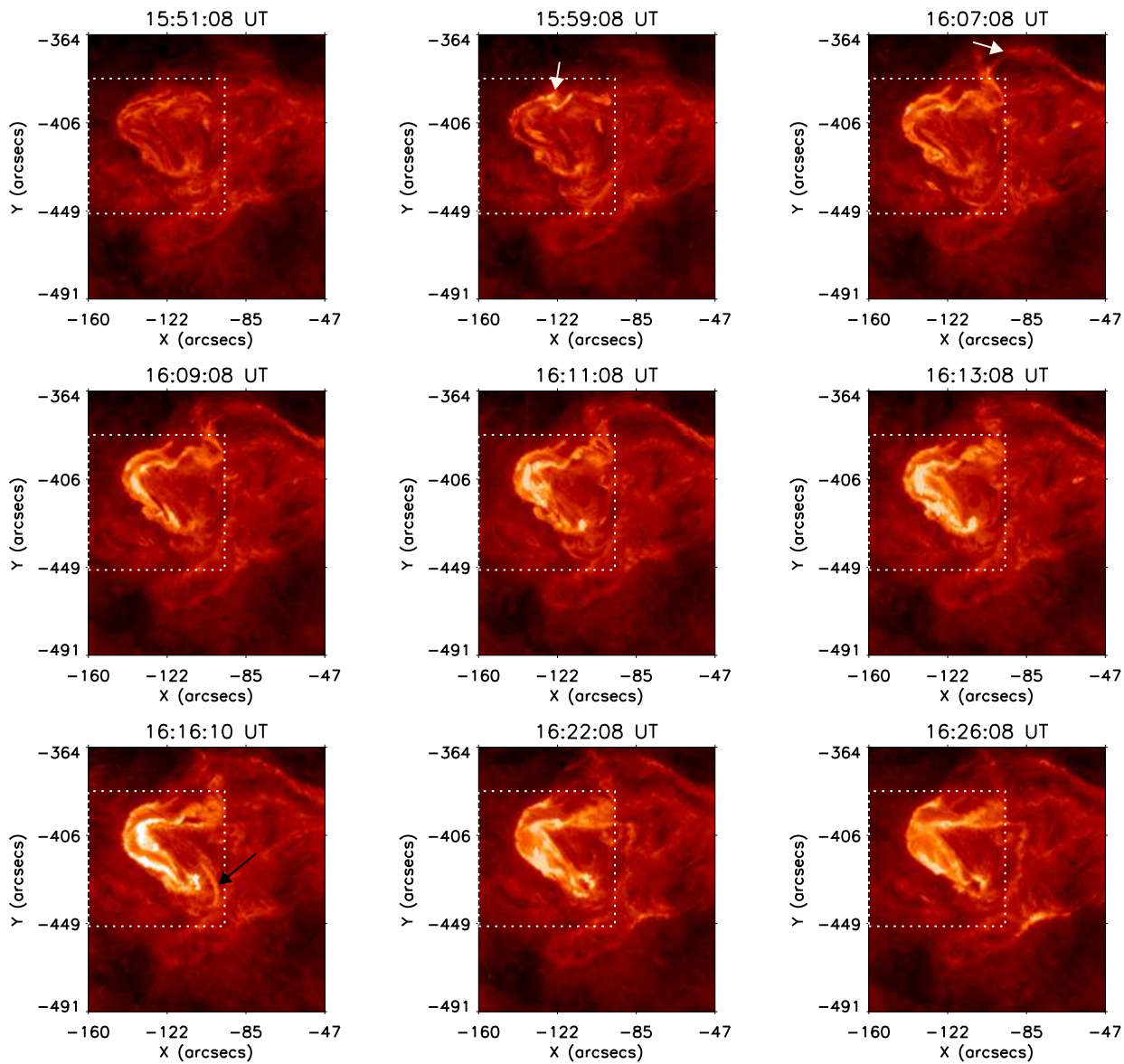}
\caption{The flare region in AIA He~{\sc ii} 304~\AA\ channel. The arrow at 15:59~UT indicates the first
coronal eruption. The arrow at 16:07\,UT denotes a rising loop. The arrow at 16:16~UT indicates the erupting filament seen in emission. The IBIS field of view
is indicted as dotted lines. Animation of the full cadence images is available on line.}
\label{fig9}
\end{figure*}



\begin{figure*}[!ht]
\centering
\includegraphics[width=17cm]{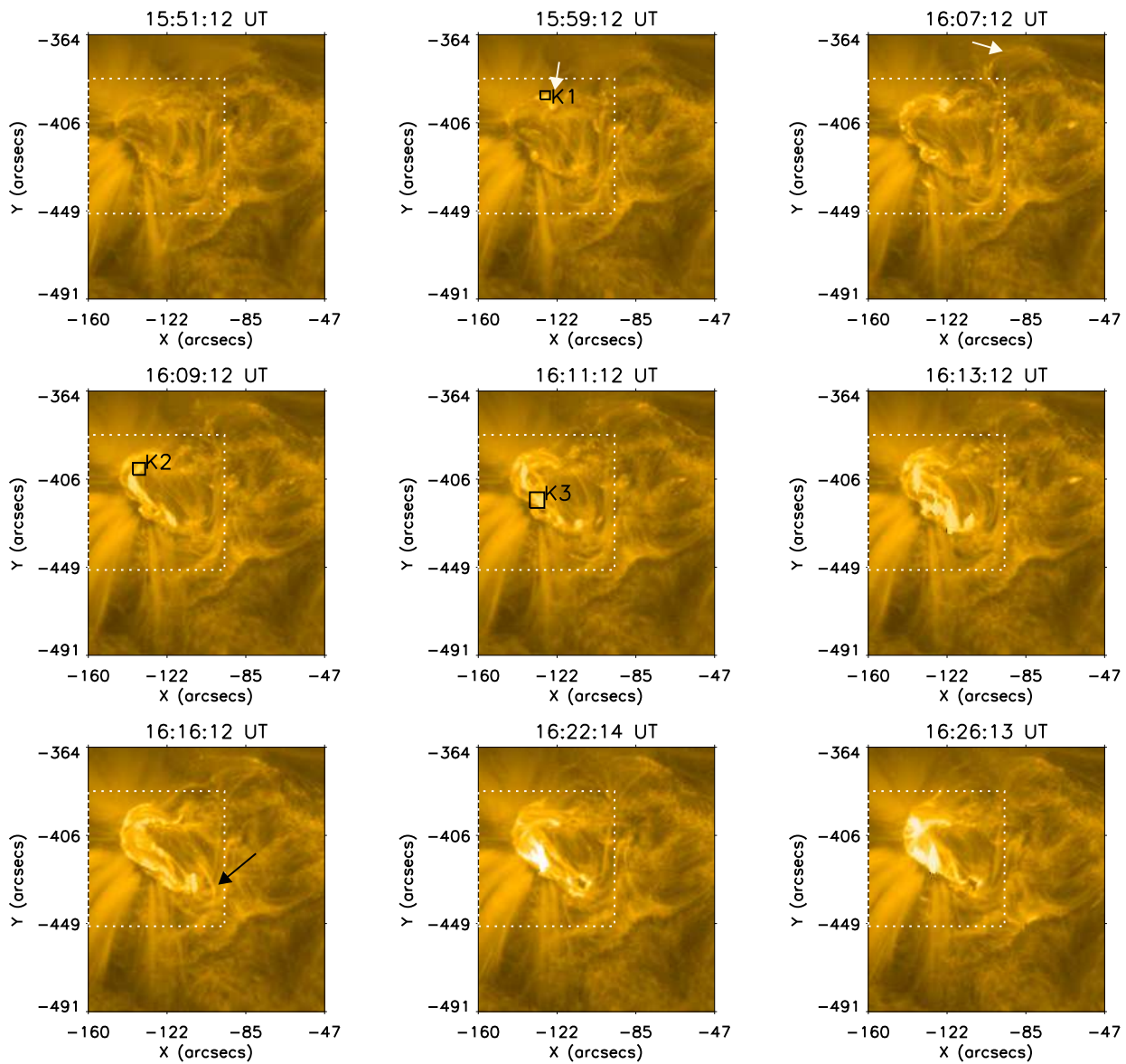}
\caption{As Fig.\,\ref{fig9}, showing the flare region seen in the AIA Fe~{\sc x} 171~\AA\ channel. The three
kernels (K1, K2, and K3; see text for details) are denoted in the images at 15:59\,UT, 16:09\,UT, and 16:11\,UT,  respectively.}
\label{fig10}
\end{figure*}


\begin{figure*}[!ht]
\centering
\includegraphics[width=17cm]{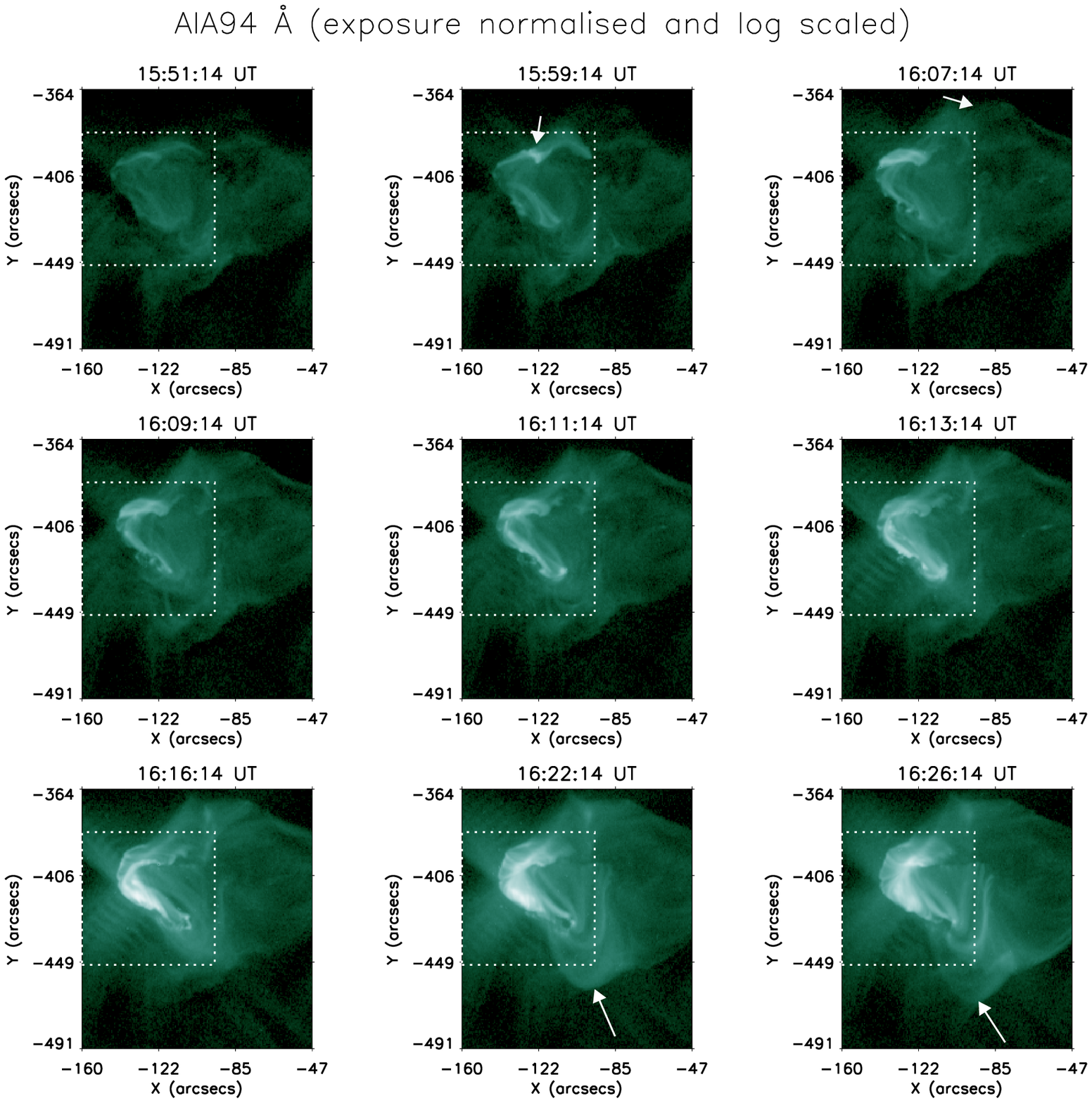}
\caption{As Fig.\,\ref{fig9}, showing the AIA Fe\,{\sc xviii}\,94~\AA\ channel. The arrows marked in the images at 16:22\,UT and 16:26\,UT show the rising coronal loops indicating the initiation of the CME.}
\label{fig11}
\end{figure*}


\section{Results and Discussion}

The combination of ground and space-based observations described above provide a nearly unique 
view of a filament eruption, flare, and CME, and the complex evolution of the solar
chromosphere, transition region, and corona during these impulsive processes. The flare that occurred during our observations was classed as GOES C4.3. 
We describe in detail the triplet of activity in the following subsections, while a general timeline can be found in Table~\ref{table1}. The spatial relation among the magnetic polarities, flare ribbons, and filament (including its legs) can be obtained from Figs.\,\ref{fig1}, \ref{fig2}, and \ref{fig3}.

\subsection{Filament eruption}
One of the main goals of the present study was to identify the possible trigger(s) and precursor(s) of
the filament destabilisation.  From the visual analysis of the speckle blue- and red-wing images
(Figs.~\ref{fig1} and \ref{fig2}) taken at \halpha$\pm$0.8~\AA\  we identified an increased activity in one
of the feet of the prominence. If we interpret the increased absorption in the wings as a Doppler velocity signature, it appears
that cool material is rising and draining along one of the filament legs (see the region noted
by a white arrow at 15:51:15~UT in Fig.~\ref{fig1}). While the up-flow lasted only approximately  5~mins, the
plasma draining persisted for almost 10~mins and was followed by a slow ascent and then an abrupt eruption of the
filament. A circle ribbon at the filament footpoint together with simultaneous rising and falling
material seen as increased absorption in the blue and red wings, strongly suggests a surge-like activity
occurring at this location. 
Unfortunately, we are not able to determine the trigger for the surge-like activity. The HMI resolution (0.505\arcsec\ pixel size) is five times lower than the IBIS resolution (0.1\arcsec) and this does not permit us to identify the small-scale magnetic flux changes in the footpoint of the surge-like phenomenon. The size of this footpoint is  
$\sim$3 arcsecs.  Magnetic flux emergence is believed to be the trigger for a surge formation \citep{2007A&A...469..331J}, but so far it has been studied only for larger events because of resolution constraints. Line-of-sight effects prevent us from establishing whether the ribbon is related to the
flows in the footpoint or if a separate surge-like event at the filament footpoint takes place. The timeline of
the events is given below.

The dynamic evolution of the region showed a sharp brightness increase in the  circular ribbon
at $\approx$15:45~UT which is also the time when the up- and down-flows of cool plasma started in the filament footpoint.  Plasma draining in
 filament footpoints (often in just one foot) is a typical precursor  for filament destabilisation (see Section\,\ref{sect_intro} for more details).  
In addition, the analysis of the total magnetic flux (integrated over the active region area outlined with a black dashed line in the left panel of Fig.~\ref{fig3}) 
in the HMI longitudinal  magnetograms clearly shows an abrupt increase in around 3\%\ of the positive flux in just 10~mins. 
The integrated flux reaches its maximum
by the time the first sign of destabilisation was recorded and then
sharply decreases during the duration of the filament-eruption and flare activity (Fig.~\ref{fig3}).
The change in the magnetic field will be discussed later in Sect.~3.4. The
filament eruption started at 15:59~UT with a transverse component of the rise velocity of 8\,km\,s$^{-1}$. During the eruption phase, the
velocity (estimated from the time-slice plots) increased from 36~\kms\ to 85~\kms\ from 16:09~UT to 16:16~UT. {We 
note that while the whole filament rose, only part of it, approximately one quarter appears to have erupted. We cannot conclude with certainty whether
 this filament eruption can be classified as partial (with some of the material that escapes with the CME) or failed. We clearly see that part of the filament falls 
back (see the white arrow in Fig.~\ref{fig2} at 16:16:09~UT pointing at the down-falling filament material, and animation Fig.\,\ref{figA4} online), but we cannot
reject the possibility that some of the prominence material made it into the coronal mass ejection
(CME; see Section~\ref{cme}).} The filament rise and fall are also seen in emission in all
AIA channels shown in Figs.~\ref{fig9}, \ref{fig10}, and \ref{fig11} as well as the animated material given online (Fig.~\ref{figA4}). The erupting filament seen in
emission in these channels suggests heating to temperatures as high as $logT(K)=5.8$. We note that the
171~\AA\ and 304~\AA\ channels have a significant contribution from cooler emission \citep[for details see][]{2011A&A...535A..46D},
which could explain the appearance of the cooler prominence material in these passbands. There is a clear indication that the erupting filament
pushed into the overlying loops which then erupted as part of the CME (see the animation online Fig.~\ref{figA4}). A 2$\pi$
twist of the filament is clearly visible while rising, best seen in
\halpha--0.8~\AA\ at 16:11~UT (Fig.~\ref{fig1}). The time-slice plots shown in Fig.~\ref{fig4} are produced by cutting through the surge-like event  to illustrate the sequence
of events described above.

\subsection{Flare evolution}
\label{sect_flare}
The multi-instrument approach of the present analysis was crucial for untangling the complexity of the
filament--flare activity. We especially took advantage of the high-cadence and high-resolution imaging and
spectroscopy of these events in order to investigate in great detail their temporal and spatial evolution at small
scale. As we discussed above, the \halpha\ observations clearly indicated that
the flare was triggered by the filament destabilisation and eruption. The GBM/Fermi lightcurves (Fig.~\ref{fig5})
 point towards three phases of the flare evolution before the decay phase, each with different energetics. We were able to determine
the events that relate to each of these phases by combining temporally and spatially the chromospheric, EUV
soft X-ray, and hard X-ray observations. We identified the formation of three flare ribbons (noted as R1, R2, and R3 in Fig.~\ref{fig2}) which occur consecutively during these phases.  All three ribbons are clearly seen in the \halpha\
blue and red wings with a stronger presence in the red wing (as discussed in detail in Section~\ref{asym}).
We selected and analysed in detail three chromospheric kernels, one from each ribbon, and discuss them in detail together with their corresponding coronal activity.
The kernels are shown with boxes in Fig.~\ref{fig2} (red wing of \halpha) and Fig.~\ref{fig10} (AIA\,171 channel) 
and are labeled as K1, K2, and K3. The kernels corresponding to the lightcurves in \halpha\ and AIA EUV 
channels are given in Figs.~\ref{fig6} and \ref{fig7}. The box sizes were chosen to include the moving kernel 
during its evolution (see Section~\ref{sect_intro} on kernel characteristics).

{\bf K1:} Kernel K1 belongs to the flare ribbon which initially appeared in the active region during the first phase (from approximately 16:00~UT to 16:09~UT). The lightcurves of the kernel in the \halpha\ wings and line centre are shown in Fig.~\ref{fig6}, top panel. Please note that the speckle images (at \halpha$\pm$0.8~\AA) were used to produce the wing lightcurves in order to take advantage of the better spatial resolution and stability of these images. The lightcurves reveal that the most significant increase in the emission is recorded in the line centre while the wing's response is much weaker and shorter in time.

\par
The intensity peak in  this kernel is at approximately 16:02~UT and lasts until 16:06~UT. It reflects the chromospheric
response to the coronal activity which started as early as 15:44:36~UT in AIA~171~\AA, 15:44:43~UT in AIA 193~\AA,
and 15:45:44~UT in AIA 304~\AA\ (see the animation online Fig.~\ref{figA4}). This flare precursor activity slows
down for a few minutes and then takes off quickly at $\sim$15:54~UT. Its location is shown with a white arrow in
Figs.~\ref{fig9} -- \ref{fig11} at 15:59~UT.  A small-brightening in the corona at the junction of two loop systems 
with different orientations can be seen at this time that produces a significant peak in the coronal
emission at around 16:02~UT (top panel of \ref{fig7}). A rising loop is seen clearly
after 16:06~UT (see the arrows at 16:07~UT in Figs.~\ref{fig9} -- \ref{fig11} and the animation in Fig.~\ref{figA4}).
The coronal activity results in a weak two-ribbon formation which can be clearly seen in the image in
Fig.~\ref{fig2} at 15:59:20~UT where kernel K1 was also found. A slight increase in the hard X-ray  emission is detected by the
GBM detectors in the energy channel from 4.2~keV to 14.6~keV shortly after 16:00~UT which lasts until around 16:09~UT.
In addition, a weak emission increase is also noticeable in the GOES soft X-ray detector in the range of
1.0 -- 8.0~\AA\  (see Fig.~\ref{fig5}).

{\bf K2:} Kernel K2 was formed as part of the ribbon that appears while the filament is rising  up during the
second phase. The corresponding lightcurves are shown in Figs.~\ref{fig6} and \ref{fig7}, middle panels. The
kernel is first seen in the line centre while the signature in the blue and red wings is recorded for
a much shorter period of time. As in K1, the increase of the \halpha\ emission in the line centre dominates the
relative intensity increase in the wings. A stronger red asymmetry is also observed (for more details on the asymmetry
see the next section). The \halpha\ emission starts  to increase sharply after 16:09~UT reaching its maximum
at 16:10~UT. The same evolution is seen in the EUV emission (Fig.~\ref{fig7}, middle panel). We note that while
the AIA 304~\AA\ and  171~\AA\ channels peak at the same time as the \halpha\ line centre, the 94~\AA\ response
is almost 3~mins later. The most plausible explanation is that this channel is dominated by emission from higher temperature plasma
(mainly from Fe~{\sc xviii} logT$\sim$6.8~K and some from Fe~{\sc xx} logT$\sim$7.0~K) while the other two channels 
record mostly cooler transition region and coronal plasma
(logT$\sim$4.7~K and 5.9~K).  The AIA 304\,\AA\ and 171\,\AA\ channels may pick up footpoint emission due to non-thermal electrons. Therefore, the response in these channels may precede that in the AIA 94\,\AA\ channel which represents emission from the loop top due to chromospheric evaporation. In Fig.~\ref{fig8} we show a series of images that clearly demonstrate how the ribbon
forms while the filament footpoint makes a lateral movement. This evolution can also be followed in the online
material (Fig.~\ref{figA4}). There is little doubt that the reconnection in the corona is triggered by the rising and
expanding filament pushing the overlying coronal structures. The timing of these events coincides clearly with
the eruption phase of the filament which starts at $\sim$16:09~UT (see Section~3.1). A second sharp increase in the hard
(GBM) and soft (GOES) X-ray emission starts just before 16:09~UT  and reaches peak emission
at $\sim$16:10~UT. During this phase, GBM/Fermi records hard X-ray emission in both channels: 4.2 keV to 14.6 keV and
14.6 keV to 24.9 keV. The next peak in the intensity curve in this kernel comes from the major flare which reaches a
maximum at 16:15~UT.

{\bf K3:} The K3 kernel (Figs.~\ref{fig6} and \ref{fig7}, bottom panels) is part of the large ribbon resulting
from the strongest energy release during the third phase. The \halpha\ and the AIA 304~\AA\ and 171~\AA\ channels
show a very sharp increase with the two AIA channels having their peak emission at 16:14~UT followed by the AIA
94~\AA\ channel almost 5 min later. One of the features seen is the dip in the \halpha\ blue-wing
lightcurve which can be explained by the erupting filament (see the blue-wing lightcurve in Fig.~\ref{fig6}, bottom panel).
The same relative intensity variations in the line \halpha\ centre and  wings as in K1 and K2 are also observed here. Both the hard and
soft X-ray emissions from GBM and GOES have the highest impulsive peak at 16:15~UT.

We also  analysed the observations of the AIA\,1600\,\AA\ channel, whose emission is dominated by continuum emission with some contribution by C~{\sc iv} lines (1548~\AA\ and 1550~\AA). Although the filament is not been seen in 1600\,\AA, the emission in the flare kernels seen in this channel shows similar evolution to that seen in the 304\,\AA\ channel.

\subsection{\halpha\ spectroscopy study (blue/red asymmetry)}
\label{asym}
Spectroscopic analysis of the solar chromosphere, for example \halpha\ line profile analysis, provides crucial
information for understanding the impact of the energy release during solar and stellar flares on the atmosphere.
Thanks to the IBIS \halpha\ line profile observations, we are able to report on this line profile evolution during
various phases of a solar flare. This can help to identify the mechanism of energy transfer towards the lower
atmosphere (i.e. the chromosphere, as discussed in the Introduction). Fig.~\ref{figA1} (online material)
presents the average \halpha\ line profiles
from a 5$\times$5\,px$^2$ (0.5\arcsec$\times$0.5\arcsec) box in the centre of kernel K1 (described in detail in
Section~\ref{sect_flare}). Consistent with the lightcurve of this kernel (see Fig.\,\ref{fig6}, top panel), the
corresponding \halpha\ line profiles show an enhancement in the line centre starting at 15:53~UT. 
Please note that the far wings
of \halpha\ are considered optically thin \citep{1984SoPh...93..105I} and the response there can be attributed
to Doppler shifts rather than to the optical thickness of the line. We found that the increase in the line centre
emission precedes  the emission increase in the line wings by 5~mins. This is also clearly visible in the line profiles
shown in Fig.~\ref{figA1}. If the emission in the line centre is assumed  to come entirely  from heating and the
response in the far line wings from Doppler shifts, the  delay would represent the time between the heating
(either from energetic particles or thermal conduction) and the chromospheric condensation down-flow registered
in the chromosphere. No blue asymmetry was detected in this ribbon. The profiles show a very minor reversal from
absorption to emission, possibly because this ribbon was too weak, suggesting low energy release. This is also
confirmed by the weak hard and soft X-ray emission  registered after 15:45~UT until 16:09~UT by GBM and GOES,
respectively (see Fig.~\ref{fig5}), defined as phase 1 of the flare evolution. As discussed by
\citet{1984ApJ...282..296C}, the \halpha\ line width is also an important indicator of the heating
mechanism during a flare (see Section~1 for details). We calculated the line width of those
\halpha\ line profiles that can be well fitted by a single Gaussian function. The results are shown
in Fig.~\ref{figA1}. The full width at half maximum\,(FWHM) of \halpha\ increases from 1.57\,\AA\ at 15:58\,UT
to 2.04\,\AA\ at 16:04\,UT, and then decreases to 1.67\,\AA\ at 16:07\,UT. However, the line width does not always increase, and it decreases at the
earlier stage, e.g. at 15:59\,UT, 16:00\,UT, and 16:01\,UT. The complex evolution of the \halpha\ line profiles suggests
that possibly several heating mechanisms are at work in this ribbon.

The evolution of the \halpha\ line profiles taken from the centre of K2 is presented in Fig.~\ref{figA2}
(online material). The emission increase in the line centre begins at 16:03~UT. The energy released in this kernel
(during the second phase of the flare)  is stronger than K1, producing clear emission profiles. The first such profile,
together with the first sign of red asymmetry, is seen at $\sim$16:08:52~UT which also coincides with the beginning
of the impulsive peak seen in the hard and soft X-ray emission (Fig~\ref{fig5}). The delay between the emission increase in the
line centre and the occurrence of the red asymmetry as determined in the far wings of \halpha\ is again $\sim$5~minutes,
as in K1. The second reversal into emission of the \halpha\ line profile appears during the strongest impulsive
peak of the flare from 16:16\,UT to 16:25\,UT. This supports  the interpretation that non-thermal electron beams
are mostly  responsible for the  chromospheric heating. Red asymmetry is clearly present in the emission line
profiles (e.g. 16:09:27\,UT, 16:09:45\,UT, 16:18:36\,UT, etc). Again, no blue asymmetry is found in this kernel.
From 16:14\,UT to 16:15\,UT, \halpha\ line profiles in this kernel turn back to absorption due to the rising
filament which crosses the area of this kernel.
From 16:03\,UT to 16:35\,UT, the line width in this kernel does not exceed 1.95\,\AA. Comparing this to
1.74\,\AA\  at 16:03\,UT when the ribbon starts flaring, the small increase in the line width suggests that
non-thermal electron beam and thermal conduction work together to heat this kernel.

Fig.~\ref{figA3} (online material) then shows the \halpha\ line profile evolution in K3 (located in the main ribbon)
during the third impulsive peak. The emission in the line centre in this kernel starts increasing at 
$\sim$16:05\,UT. The first  emission profile is recorded shortly before 16:14 and lasts until 16:23\,UT, i.e. 
9\,mins. The red asymmetry appears from 16:11\,UT to 16:17\,UT. Thus, the delay between the line centre emission and the red asymmetry is $\sim$6~mins. Blue asymmetry is observed after 16:30\,UT. However, we find that it is not associated with the ribbon but 
with the red-wing absorption from the down-falling filament material.

In order to study the asymmetry in the \halpha\ line profiles over the whole flaring region, we developed
an algorithm to automatically detect these phenomena by comparing the total intensity of three points of
wavelength in the blue and red wings of the line profiles. The difference between the total intensity
of the blue and red wings is then calculated. Differences that exceed three sigma are considered to be a
wing asymmetry. Furthermore, we randomly selected more than one hundred samples in the flare ribbon and visually examined
their line profiles in order to confirm that the detection from the automatic procedure is reliable.
 We found that blue asymmetry is only associated with the dynamics of the filament,
and no blue asymmetry is identified in the ribbons where only red asymmetry is found. We note that significant red
asymmetry is also associated with the filament  plasmas. This result is consistent with
\citet{1990ApJ...363..318C} and supports their suggestion for a strong connection between a \halpha\ blue asymmetry
and dynamic filaments.

\subsection{Magnetic field evolution}
We used HMI/SDO line-of-sight magnetograms to study the magnetic field evolution before and during the flare. The
left panel of Fig.~\ref{fig3} shows a HMI magnetogram at 16:14\,UT. As mentioned above, the filament destabilisation
is possibly due to the new positive flux emergence. Unfortunately,
the resolution and sensitivity of the HMI magnetograms are not sufficient to establish with enough certainty where the crucial
emergence has taken place. Nevertheless, the increased activity in one of the filament footpoints (see Section~3.1)
 leads to the filament destabilisation and following eruption. The lightcurves of the positive and the negative
flux reveal a very high rate of magnetic flux cancellation of $1.34\times10^{16}$\,Mx\,s$^{-1}$
\citep[see the definition of cancellation rate in][]{2012A&A...548A..62H}.
This cancellation rate is two orders of magnitudes larger than that found in \citet{2012A&A...548A..62H} for an
X-ray jet, and is consistent with their suggestion that a more explosive event in the solar atmosphere is
likely to be associated with a larger magnetic cancellation rate. This also suggests that flux cancellation might drive both small- and large-scale explosive events in the solar atmosphere.

\begin{figure*}[!ht]
\centering
\includegraphics[trim=0mm 0mm 15mm 20mm,clip]{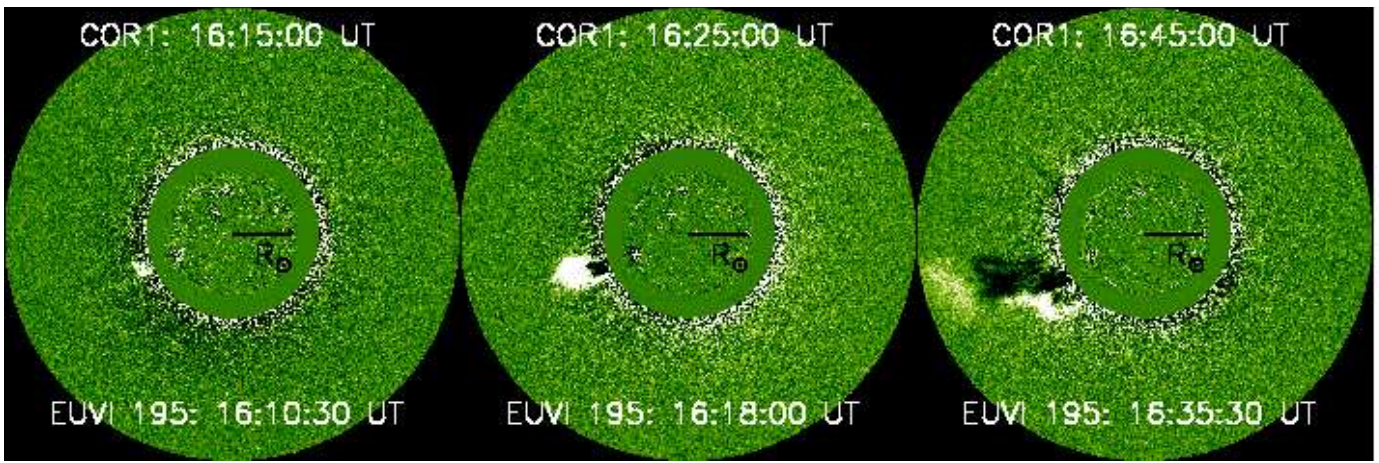}
\caption{Combination of SECCHI/COR1 and EUVI 195~\AA\ Ahead images of the CME caused by the filament/flare sequence of events. }
\label{fig12}
\end{figure*}

\subsection{CME association}
\label{cme}

The CME initiation was recorded by both EUVI instruments on STEREO A and B thanks to the suitable position angle of the two satellites. By using an image difference technique, we analysed the CME lift-off in the EUVI A and B images. The signature of the ejected material is very faint but, nevertheless, clear enough to identify the erupting material.
A coronal mass ejection was registered by LASCO C2 on board SoHO, and by COR1 and COR2 on board the  STEREO Ahead and Behind spacecrafts. The CME was first registered in SECCHI COR1 at 16:15~UT (Fig.~\ref{fig12}). It  propagated with an average speed of 367~\kms (minimum 250~\kms\ and maximum 595~\kms)   as measured in SECCHI COR2 (from CACTus automatic measurements\footnote{http://sidc.oma.be/cactus/}).  We measured an angular width of the CME of 59$^{\circ}$ determined in SECCHI COR2 (CACTus gives 66$^{\circ}$).   Two clouds are seen forming the CME, with the first one appearing at approximately 16:15~UT followed by the second one displaced more to the south between 16:25~UT and 16:30~UT.  The two parts of the CME reflect the complexity of the flare event described above. We can speculate that the two separate clouds ejected with a time difference of $\sim$10--15~mins (note the 5 min cadence of the images) coincides with the two strong impulsive peaks recorded during the flare. The first cloud initiats in the region above kernel K1 while the second forms during the next two phases.

\section{Summary}
\label{sum}
In the present study we analysed multi-instrument observations of a GOES C4.3 solar flare with filament
destabilisation and eruption that occurred in  NOAA 11123 on 2010 November 11. As a result of this activity a
coronal mass ejection formed. We used a unique combination of ground- and space-based spectral and imaging observations 
 to explore in great detail both temporally and spatially  the small-scale evolution of all three
phenomena. The main results can be summarised as follows:

We found that the filament destabilisation and eruption was the main trigger for the flaring activity.  A surge-like
event with a circular ribbon at the filament footpoint was determined to be the trigger of  the filament
destabilisation. Draining in one of the footpoints of the filament was identified as the precursor for the eruption.

HMI line-of-sight magnetograms revealed magnetic flux emergence prior to the filament destabilisation
followed by a high rate of flux cancellation of $1.34\times10^{16}$\,Mx\,s$^{-1}$ during the flare activity.

GBM/Fermi and GOES  lightcurves reveal a complex flare evolution with three phases with different energetics.
Each phase was associated with a particular chromospheric (three separate ribbon formations) and coronal
activity (brightenings and loop ejections). Three kernels were studied in detail for each ribbon. In each kernel
the intensity increase in the \halpha\ line centre is much stronger than the emission increase in the line wings.
This is consistent with the latest simulations by \citet{2009A&A...499..923K} showing that non-thermal electron
beams result in emission enhancement and have only a small contribution to the Doppler wings. In all three kernels,
the delay between the line centre emission increase and the red asymmetry was around 5--6~mins. We have to
point out that the interpretation of the \halpha\ line is not straightforward because of the optical thickness and complex formation of this
line.

No reversal to emission in the \halpha\ was found during the weakest phase of the flare, which is consistent with
the hard X-ray emission which registered the event only in the 4.2 - 14.6 keV range. A second phase is related to
the first impulsive peak registered in hard (up to 25 keV) and soft X-rays. During this phase \halpha\ line
emission profiles were observed. The strongest impulsive peak produced hard X-ray emission in the range of 25 -- 50 keV,
but most of the hard X-ray flux comes from the lowest energy band of 4.2 to 15~keV. In the flare ribbons, only red asymmetry 
is found during all three phases. In contrast, the  blue asymmetry
is entirely associated with the dynamic filament.

The filament-eruption/flare event  produces  a narrow coronal mass ejection comprising  two clouds
released with a time difference of approximately 10 min which reflects the complexity of the energy release
during this flare. The filament displays a failed eruption, but presumably some of the filament material
escaped and so could not be detected in the fields of view or temperature ranges of the available instrumentation. While the whole filament appears to have been destabilised 
only part of it had erupted.

The data presented in this study will be further explored both for comparison with data driven three-dimensional magneto-hydrodynamic  numerical simulations as well as \halpha\ line profile calculations. A separate study will analyse in more detail  the coronal mass ejection detected during this event.

\begin{acknowledgements}
This research is supported by the China 973 program 2012CB825601, and the National Natural Science Foundation of China under contracts: 41274178, 40904047, 41174154, 41274176. Research at the Armagh Observatory is grant-aided by the N. Ireland Dept. of Culture, Arts and Leisure. We thank PATT T\&S for their support, and are grateful for support from STFC grant ST/J001082/1 and the Leverhulme Trust. We thank the National Solar Observatory Sacramento Peak for their hospitality in particular Doug Gilliam for his help during the observations. We thank Dr. Gianna Cauzzi for her great help with the IBIS data reduction.  We would also like to thank Friedrich W\"oger for his KISIP code. We also thank very much Kim Tolbert for her great support for handling the Fermi data. We thank Klaus Galsgaard for the careful reading and constructive suggestions on the manuscript. AIA and HMI data is courtesy of SDO (NASA). We would like to thank the anonymous referee for his/her critical and constructive comments. We thank Marilena Mierla for the helpful discussion on the coronal mass ejection.
\end{acknowledgements}

\bibliographystyle{aa}
\bibliography{references1}

\begin{thebibliography}{57}
\expandafter\ifx\csname natexlab\endcsname\relax\def\natexlab#1{#1}\fi

\bibitem[{{Alexander} {et~al.}(2006){Alexander}, {Liu}, \&
  {Gilbert}}]{2006ApJ...653..719A}
{Alexander}, D., {Liu}, R., \& {Gilbert}, H.~R. 2006, \apj, 653, 719

\bibitem[{{Allred} {et~al.}(2005){Allred}, {Hawley}, {Abbett}, \&
  {Carlsson}}]{2005ApJ...630..573A}
{Allred}, J.~C., {Hawley}, S.~L., {Abbett}, W.~P., \& {Carlsson}, M. 2005,
  \apj, 630, 573

\bibitem[{{Asai} {et~al.}(2012){Asai}, {Ichimoto}, {Kita}, {Kurokawa}, \&
  {Shibata}}]{2012PASJ...64...20A}
{Asai}, A., {Ichimoto}, K., {Kita}, R., {Kurokawa}, H., \& {Shibata}, K. 2012,
  \pasj, 64, 20

\bibitem[{Benz(2008)}]{lrsp-2008-1}
Benz, A.~O. 2008, Living Reviews in Solar Physics, 5

\bibitem[{{Canfield} {et~al.}(1984){Canfield}, {Gunkler}, \&
  {Ricchiazzi}}]{1984ApJ...282..296C}
{Canfield}, R.~C., {Gunkler}, T.~A., \& {Ricchiazzi}, P.~J. 1984, \apj, 282,
  296

\bibitem[{{Canfield} {et~al.}(1990{\natexlab{a}}){Canfield}, {Metcalf},
  {Zarro}, \& {Lemen}}]{1990ApJ...348..333C}
{Canfield}, R.~C., {Metcalf}, T.~R., {Zarro}, D.~M., \& {Lemen}, J.~R.
  1990{\natexlab{a}}, \apj, 348, 333

\bibitem[{{Canfield} {et~al.}(1990{\natexlab{b}}){Canfield}, {Penn}, {Wulser},
  \& {Kiplinger}}]{1990ApJ...363..318C}
{Canfield}, R.~C., {Penn}, M.~J., {Wulser}, J.-P., \& {Kiplinger}, A.~L.
  1990{\natexlab{b}}, \apj, 363, 318

\bibitem[{{Canfield} \& {Reardon}(1998)}]{1998SoPh..182..145C}
{Canfield}, R.~C. \& {Reardon}, K.~P. 1998, \solphys, 182, 145

\bibitem[{{Carmichael}(1964)}]{1964NASSP..50..451C}
{Carmichael}, H. 1964, NASA Special Publication, 50, 451

\bibitem[{{Cauzzi} {et~al.}(2009){Cauzzi}, {Reardon}, {Rutten}, {Tritschler},
  \& {Uitenbroek}}]{2009A&A...503..577C}
{Cauzzi}, G., {Reardon}, K., {Rutten}, R.~J., {Tritschler}, A., \&
  {Uitenbroek}, H. 2009, \aap, 503, 577

\bibitem[{{Cavallini}(2006)}]{2006SoPh..236..415C}
{Cavallini}, F. 2006, \solphys, 236, 415

\bibitem[{{Chen}(2011)}]{2011LRSP....8....1C}
{Chen}, P.~F. 2011, Living Reviews in Solar Physics, 8, 1

\bibitem[{{Cheng} {et~al.}(2006){Cheng}, {Ding}, \& {Li}}]{2006ApJ...653..733C}
{Cheng}, J.~X., {Ding}, M.~D., \& {Li}, J.~P. 2006, \apj, 653, 733

\bibitem[{{Cirigliano} {et~al.}(2004){Cirigliano}, {Vial}, \&
  {Rovira}}]{2004SoPh..223...95C}
{Cirigliano}, D., {Vial}, J.-C., \& {Rovira}, M. 2004, \solphys, 223, 95

\bibitem[{{Del Zanna} {et~al.}(2011){Del Zanna}, {O'Dwyer}, \&
  {Mason}}]{2011A&A...535A..46D}
{Del Zanna}, G., {O'Dwyer}, B., \& {Mason}, H.~E. 2011, \aap, 535, A46

\bibitem[{{Deng} {et~al.}(2013){Deng}, {Tritschler}, {Jing}, {Chen}, {Liu},
  {Reardon}, {Denker}, {Xu}, \& {Wang}}]{deng2012}
{Deng}, N., {Tritschler}, A., {Jing}, J., {et~al.} 2013, \apj, 769, 112

\bibitem[{{Engvold} {et~al.}(2001){Engvold}, {Jakobsson}, {Tandberg-Hanssen},
  {Gurman}, \& {Moses}}]{2001SoPh..202..293E}
{Engvold}, O., {Jakobsson}, H., {Tandberg-Hanssen}, E., {Gurman}, J.~B., \&
  {Moses}, D. 2001, \solphys, 202, 293

\bibitem[{{Fang} {et~al.}(1993){Fang}, {Henoux}, \&
  {Gan}}]{1993A&A...274..917F}
{Fang}, C., {Henoux}, J.~C., \& {Gan}, W.~Q. 1993, \aap, 274, 917

\bibitem[{{Feynman} \& {Martin}(1995)}]{1995JGR...100.3355F}
{Feynman}, J. \& {Martin}, S.~F. 1995, \jgr, 100, 3355

\bibitem[{{Fletcher} {et~al.}(2011){Fletcher}, {Dennis}, {Hudson}, {Krucker},
  {Phillips}, {Veronig}, {Battaglia}, {Bone}, {Caspi}, {Chen}, {Gallagher},
  {Grigis}, {Ji}, {Liu}, {Milligan}, \& {Temmer}}]{2011SSRv..159...19F}
{Fletcher}, L., {Dennis}, B.~R., {Hudson}, H.~S., {et~al.} 2011, \ssr, 159, 19

\bibitem[{{Fletcher} \& {Hudson}(2008)}]{2008ApJ...675.1645F}
{Fletcher}, L. \& {Hudson}, H.~S. 2008, \apj, 675, 1645

\bibitem[{{Gibson} {et~al.}(2006){Gibson}, {Foster}, {Burkepile}, {de Toma}, \&
  {Stanger}}]{2006ApJ...641..590G}
{Gibson}, S.~E., {Foster}, D., {Burkepile}, J., {de Toma}, G., \& {Stanger}, A.
  2006, \apj, 641, 590

\bibitem[{{Gilbert} {et~al.}(2000){Gilbert}, {Holzer}, {Burkepile}, \&
  {Hundhausen}}]{2000ApJ...537..503G}
{Gilbert}, H.~R., {Holzer}, T.~E., {Burkepile}, J.~T., \& {Hundhausen}, A.~J.
  2000, \apj, 537, 503

\bibitem[{{Graeter} \& {Kucera}(1992)}]{1992SoPh..141...91G}
{Graeter}, M. \& {Kucera}, T.~A. 1992, \solphys, 141, 91

\bibitem[{{Harrison} {et~al.}(1985){Harrison}, {Waggett}, {Bentley},
  {Phillips}, {Bruner}, {Dryer}, \& {Simnett}}]{1985SoPh...97..387H}
{Harrison}, R.~A., {Waggett}, P.~W., {Bentley}, R.~D., {et~al.} 1985, \solphys,
  97, 387

\bibitem[{{Heinzel} {et~al.}(1994){Heinzel}, {Karlicky}, {Kotrc}, \&
  {Svestka}}]{1994SoPh..152..393H}
{Heinzel}, P., {Karlicky}, M., {Kotrc}, P., \& {Svestka}, Z. 1994, \solphys,
  152, 393

\bibitem[{{Hirayama}(1974)}]{1974SoPh...34..323H}
{Hirayama}, T. 1974, \solphys, 34, 323

\bibitem[{{Howard} {et~al.}(2008){Howard}, {Moses}, {Vourlidas}, {Newmark},
  {Socker}, {Plunkett}, {Korendyke}, {Cook}, {Hurley}, {Davila}, {Thompson},
  {St Cyr}, {Mentzell}, {Mehalick}, {Lemen}, {Wuelser}, {Duncan}, {Tarbell},
  {Wolfson}, {Moore}, {Harrison}, {Waltham}, {Lang}, {Davis}, {Eyles},
  {Mapson-Menard}, {Simnett}, {Halain}, {Defise}, {Mazy}, {Rochus}, {Mercier},
  {Ravet}, {Delmotte}, {Auchere}, {Delaboudiniere}, {Bothmer}, {Deutsch},
  {Wang}, {Rich}, {Cooper}, {Stephens}, {Maahs}, {Baugh}, {McMullin}, \&
  {Carter}}]{2008SSRv..136...67H}
{Howard}, R.~A., {Moses}, J.~D., {Vourlidas}, A., {et~al.} 2008, \ssr, 136, 67

\bibitem[{{Huang} {et~al.}(2012){Huang}, {Madjarska}, {Doyle}, \&
  {Lamb}}]{2012A&A...548A..62H}
{Huang}, Z., {Madjarska}, M.~S., {Doyle}, J.~G., \& {Lamb}, D.~A. 2012, \aap,
  548, A62

\bibitem[{{Hudson}(2007)}]{2007ASPC..368..365H}
{Hudson}, H.~S. 2007, in Astronomical Society of the Pacific Conference Series,
  Vol. 368, The Physics of Chromospheric Plasmas, ed. P.~{Heinzel},
  I.~{Dorotovi{\v c}}, \& R.~J. {Rutten}, 365

\bibitem[{{Ichimoto} \& {Kurokawa}(1984)}]{1984SoPh...93..105I}
{Ichimoto}, K. \& {Kurokawa}, H. 1984, \solphys, 93, 105

\bibitem[{{Ji} {et~al.}(1994){Ji}, {Kurokawa}, {Fang}, \&
  {Huang}}]{1994SoPh..149..195J}
{Ji}, G.~P., {Kurokawa}, H., {Fang}, C., \& {Huang}, Y.~R. 1994, \solphys, 149,
  195

\bibitem[{{Jiang} {et~al.}(2007){Jiang}, {Chen}, {Li}, {Shen}, \&
  {Yang}}]{2007A&A...469..331J}
{Jiang}, Y.~C., {Chen}, H.~D., {Li}, K.~J., {Shen}, Y.~D., \& {Yang}, L.~H.
  2007, \aap, 469, 331

\bibitem[{{Kaiser} {et~al.}(2008){Kaiser}, {Kucera}, {Davila}, {St.~Cyr},
  {Guhathakurta}, \& {Christian}}]{2008SSRv..136....5K}
{Kaiser}, M.~L., {Kucera}, T.~A., {Davila}, J.~M., {et~al.} 2008, \ssr, 136, 5

\bibitem[{{Ka{\v s}parov{\'a}} {et~al.}(2009){Ka{\v s}parov{\'a}}, {Varady},
  {Heinzel}, {Karlick{\'y}}, \& {Moravec}}]{2009A&A...499..923K}
{Ka{\v s}parov{\'a}}, J., {Varady}, M., {Heinzel}, P., {Karlick{\'y}}, M., \&
  {Moravec}, Z. 2009, \aap, 499, 923

\bibitem[{{Kopp} \& {Pneuman}(1976)}]{1976SoPh...50...85K}
{Kopp}, R.~A. \& {Pneuman}, G.~W. 1976, \solphys, 50, 85

\bibitem[{{Kucera} \& {Landi}(2006)}]{2006ApJ...645.1525K}
{Kucera}, T.~A. \& {Landi}, E. 2006, \apj, 645, 1525

\bibitem[{{Kucera} \& {Landi}(2008)}]{2008ApJ...673..611K}
{Kucera}, T.~A. \& {Landi}, E. 2008, \apj, 673, 611

\bibitem[{{Lemen} {et~al.}(2012){Lemen}, {Title}, {Akin}, {Boerner}, {Chou},
  {Drake}, {Duncan}, {Edwards}, {Friedlaender}, {Heyman}, {Hurlburt}, {Katz},
  {Kushner}, {Levay}, {Lindgren}, {Mathur}, {McFeaters}, {Mitchell}, {Rehse},
  {Schrijver}, {Springer}, {Stern}, {Tarbell}, {Wuelser}, {Wolfson}, {Yanari},
  {Bookbinder}, {Cheimets}, {Caldwell}, {Deluca}, {Gates}, {Golub}, {Park},
  {Podgorski}, {Bush}, {Scherrer}, {Gummin}, {Smith}, {Auker}, {Jerram},
  {Pool}, {Soufli}, {Windt}, {Beardsley}, {Clapp}, {Lang}, \&
  {Waltham}}]{2012SoPh..275...17L}
{Lemen}, J.~R., {Title}, A.~M., {Akin}, D.~J., {et~al.} 2012, \solphys, 275, 17

\bibitem[{{Liu} {et~al.}(2012){Liu}, {Hoeksema}, {Scherrer}, {Schou},
  {Couvidat}, {Bush}, {Duvall}, {Hayashi}, {Sun}, \&
  {Zhao}}]{2012SoPh..279..295L}
{Liu}, Y., {Hoeksema}, J.~T., {Scherrer}, P.~H., {et~al.} 2012, \solphys, 279,
  295

\bibitem[{{Magara} {et~al.}(1996){Magara}, {Mineshige}, {Yokoyama}, \&
  {Shibata}}]{1996ApJ...466.1054M}
{Magara}, T., {Mineshige}, S., {Yokoyama}, T., \& {Shibata}, K. 1996, \apj,
  466, 1054

\bibitem[{{Martin}(1980)}]{1980SoPh...68..217M}
{Martin}, S.~F. 1980, \solphys, 68, 217

\bibitem[{{Meegan} {et~al.}(2009){Meegan}, {Lichti}, {Bhat}, {Bissaldi},
  {Briggs}, {Connaughton}, {Diehl}, {Fishman}, {Greiner}, {Hoover}, {van der
  Horst}, {von Kienlin}, {Kippen}, {Kouveliotou}, {McBreen}, {Paciesas},
  {Preece}, {Steinle}, {Wallace}, {Wilson}, \&
  {Wilson-Hodge}}]{2009ApJ...702..791M}
{Meegan}, C., {Lichti}, G., {Bhat}, P.~N., {et~al.} 2009, \apj, 702, 791

\bibitem[{{O'Dwyer} {et~al.}(2010){O'Dwyer}, {Del Zanna}, {Mason}, {Weber}, \&
  {Tripathi}}]{2010A&A...521A..21O}
{O'Dwyer}, B., {Del Zanna}, G., {Mason}, H.~E., {Weber}, M.~A., \& {Tripathi},
  D. 2010, \aap, 521, A21

\bibitem[{{Pesnell} {et~al.}(2012){Pesnell}, {Thompson}, \&
  {Chamberlin}}]{2012SoPh..275....3P}
{Pesnell}, W.~D., {Thompson}, B.~J., \& {Chamberlin}, P.~C. 2012, \solphys,
  275, 3

\bibitem[{{Reardon} \& {Cavallini}(2008)}]{2008A&A...481..897R}
{Reardon}, K.~P. \& {Cavallini}, F. 2008, \aap, 481, 897

\bibitem[{{Russell} \& {Fletcher}(2013)}]{2013ApJ...765...81R}
{Russell}, A.~J.~B. \& {Fletcher}, L. 2013, \apj, 765, 81

\bibitem[{{Saint-Hilaire} \& {Benz}(2005)}]{2005A&A...435..743S}
{Saint-Hilaire}, P. \& {Benz}, A.~O. 2005, \aap, 435, 743

\bibitem[{{Schou} {et~al.}(2012){Schou}, {Scherrer}, {Bush}, {Wachter},
  {Couvidat}, {Rabello-Soares}, {Bogart}, {Hoeksema}, {Liu}, {Duvall}, {Akin},
  {Allard}, {Miles}, {Rairden}, {Shine}, {Tarbell}, {Title}, {Wolfson},
  {Elmore}, {Norton}, \& {Tomczyk}}]{2012SoPh..275..229S}
{Schou}, J., {Scherrer}, P.~H., {Bush}, R.~I., {et~al.} 2012, \solphys, 275,
  229

\bibitem[{Shibata \& Magara(2011)}]{lrsp-2011-6}
Shibata, K. \& Magara, T. 2011, Living Reviews in Solar Physics, 8

\bibitem[{{Sturrock}(1966)}]{1966Natur.211..695S}
{Sturrock}, P.~A. 1966, \nat, 211, 695

\bibitem[{{Tandberg-Hanssen}(1995)}]{1995ASSL..199.....T}
{Tandberg-Hanssen}, E., ed. 1995, Astrophysics and Space Science Library, Vol.
  199, {The nature of solar prominences}

\bibitem[{{Tang}(1983)}]{1983SoPh...83...15T}
{Tang}, F. 1983, \solphys, 83, 15

\bibitem[{{{\v S}vestka}(1976)}]{svestka1976}
{{\v S}vestka}, Z. 1976, {Solar Flare} (Springer)

\bibitem[{{{\v S}vestka} {et~al.}(1962){{\v S}vestka}, {Kopeck{\'y}}, \&
  {Blaha}}]{1962BAICz..13...37S}
{{\v S}vestka}, Z., {Kopeck{\'y}}, M., \& {Blaha}, M. 1962, Bulletin of the
  Astronomical Institutes of Czechoslovakia, 13, 37

\bibitem[{{W{\"o}ger} {et~al.}(2008){W{\"o}ger}, {von der L{\"u}he}, \&
  {Reardon}}]{2008A&A...488..375W}
{W{\"o}ger}, F., {von der L{\"u}he}, O., \& {Reardon}, K. 2008, \aap, 488, 375

\bibitem[{{Zarro} {et~al.}(1988){Zarro}, {Canfield}, {Metcalf}, \&
  {Strong}}]{1988ApJ...324..582Z}
{Zarro}, D.~M., {Canfield}, R.~C., {Metcalf}, T.~R., \& {Strong}, K.~T. 1988,
  \apj, 324, 582

\end{thebibliography}
\begin{appendix}

\section{Online material}

\begin{figure*}[!ht]
\centering
\includegraphics[trim=5mm 10mm 3mm 5mm,clip,width=\textwidth]{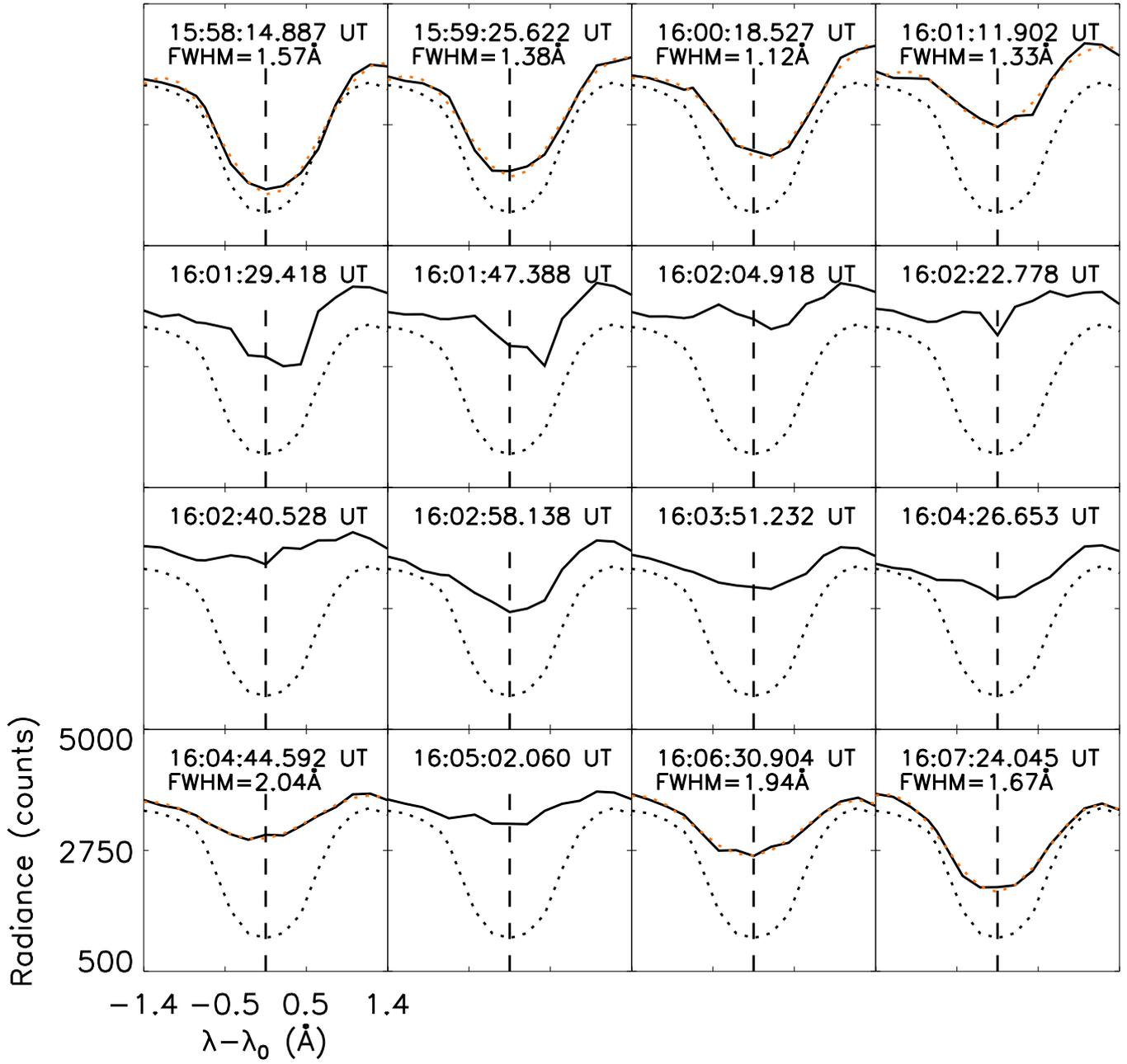}
\caption{\halpha\ line profiles (solid lines) from the kernel K1 together with a reference profile (black dotted
lines) taken from a quiet region at the bottom-left corner of the field of view. The line centre $\lambda_0$ is
denoted by dashed lines. FWHM of those profiles that can fitted by a single-Gaussian function are noted accordingly,
and their fitting Gaussian profiles (orange dotted lines) are overplotted.}
\label{figA1}
\end{figure*}

\begin{figure*}[!ht]
\centering
\includegraphics[trim=10mm 10mm 3mm 5mm,clip,width=\textwidth]{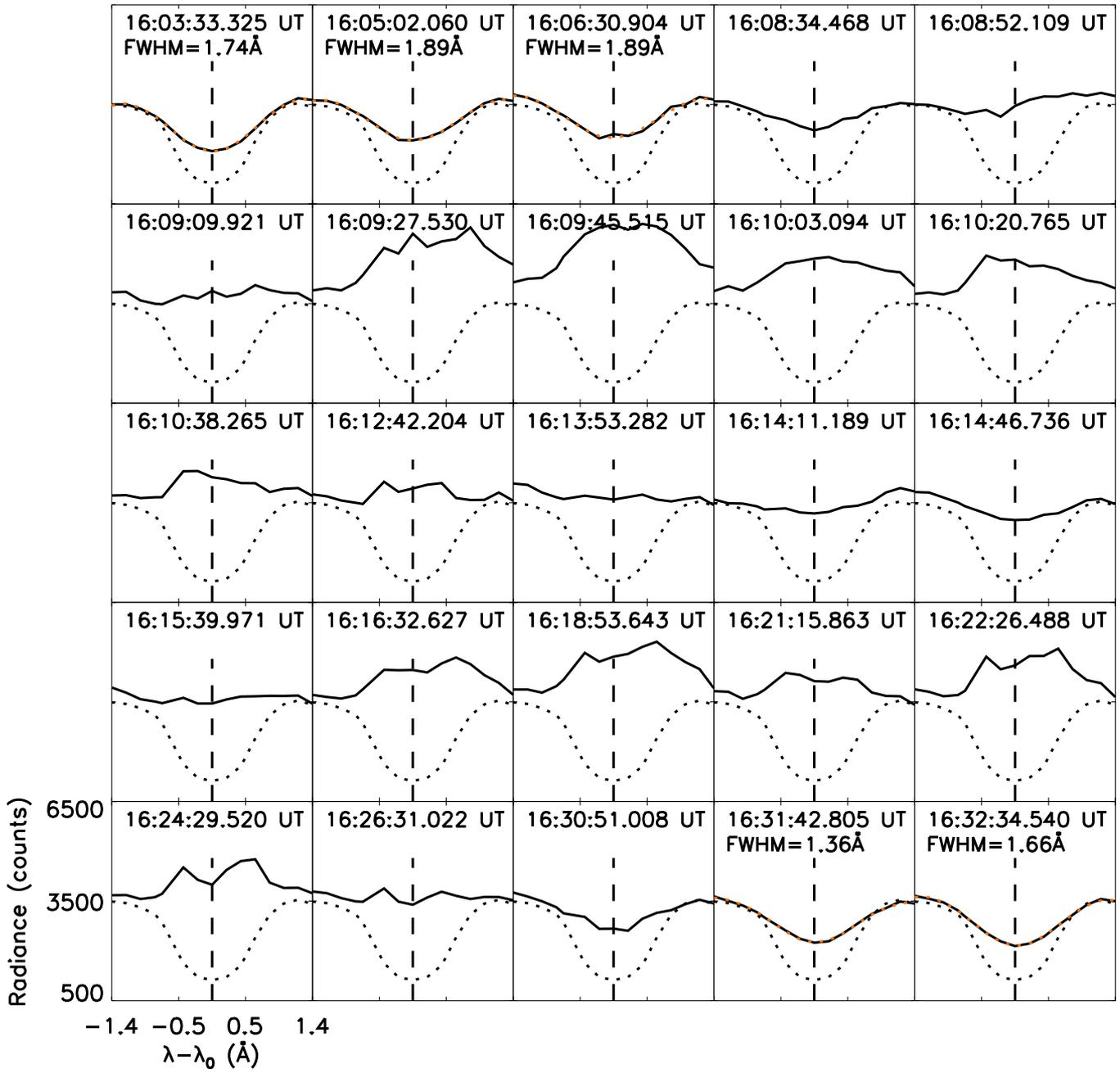}
\caption{As Fig.\,\ref{figA1}, the evolution of \halpha\ line profiles from the kernel K2.}
\label{figA2}
\end{figure*}

\begin{figure*}[!ht]
\centering
\includegraphics[trim=10mm 10mm 3mm 5mm,clip,width=\textwidth]{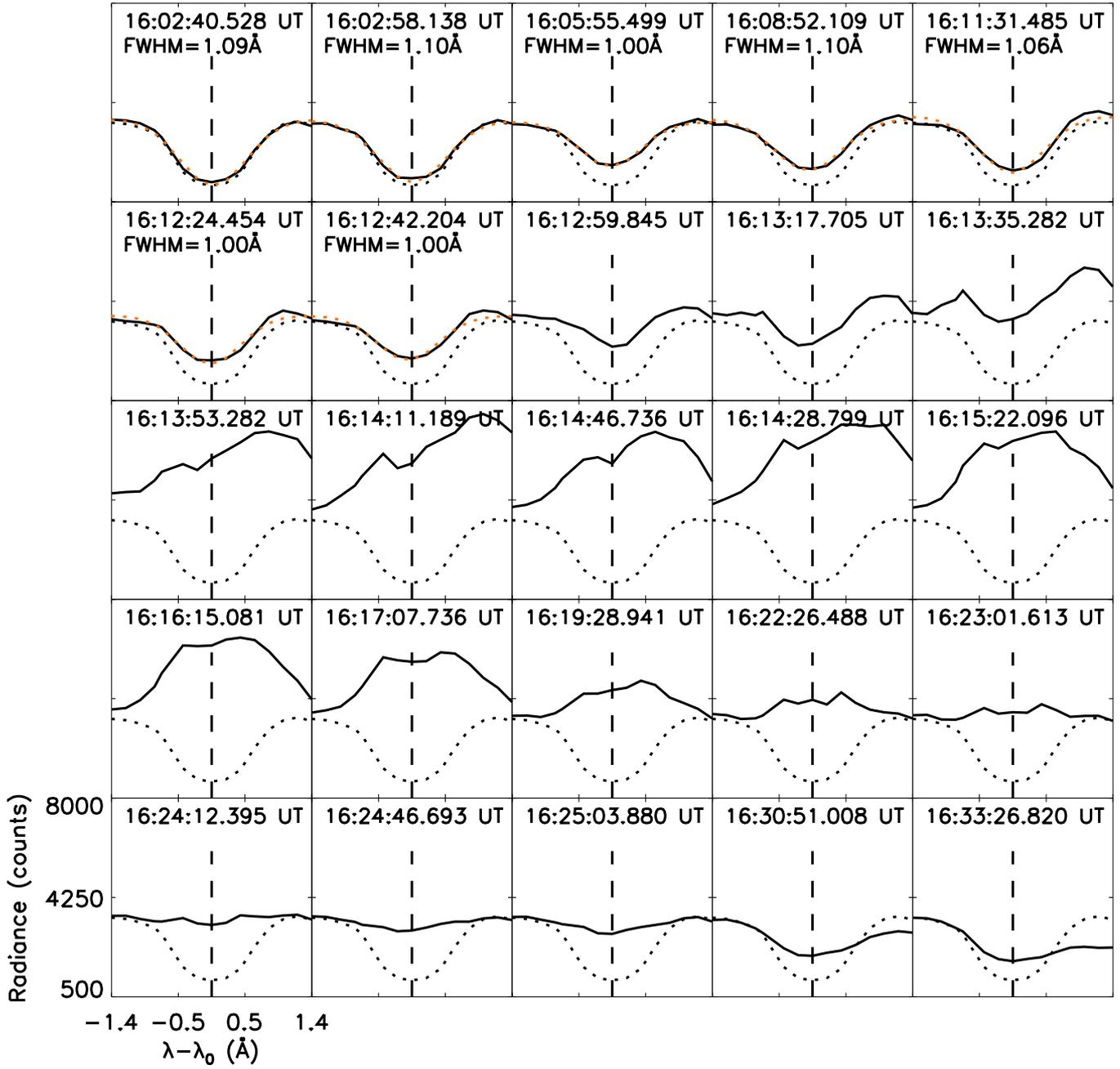}
\caption{As Fig.\,\ref{figA1}, the evolution of \halpha\ line profiles from the kernel K3.}
\label{figA3}
\end{figure*}

\begin{figure*}[!ht]
\centering
\includegraphics[width=\textwidth]{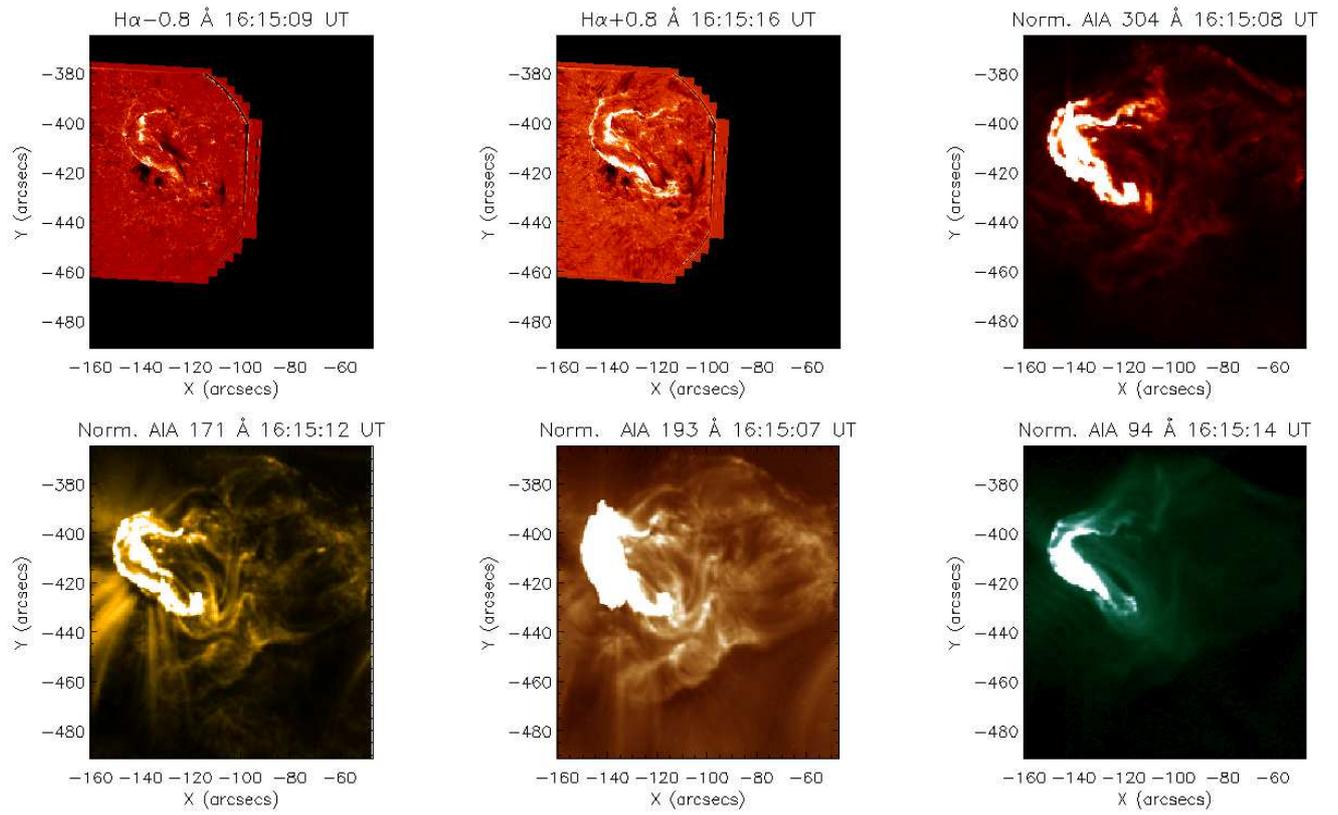}
\caption{An animation displaying the evolution of the flare seen in six different channels. Top-left: \halpha\ -- 0.8\,\AA; top-middle: \halpha\ + 0.8\,\AA; top-right: AIA 304\,\AA; bottom-left: AIA 171\,\AA; bottom-middle: AIA 193\,\AA; bottom-right: AIA 94\,\AA.}
\label{figA4}
\end{figure*}

\end{appendix}

\end{document}